\documentclass[aps,pra,reprint,amsmath,amssymb,superscriptaddress,longbibliography]{revtex4-1}

\usepackage[utf8]{inputenc}
\usepackage{graphicx,hyperref,siunitx,textcomp,soul}
\usepackage[all]{hypcap}
\graphicspath{{img/}}
\hypersetup{
  colorlinks,%
  citecolor=blue,%
  filecolor=blue,%
  linkcolor=blue,%
  urlcolor=blue
}

\newcommand*{\bra}[1]{\ensuremath{\langle #1 \vert}}
\newcommand*{\ket}[1]{\ensuremath{\vert #1 \rangle}}
\newcommand*{\inner}[2]{\langle #1 | #2 \rangle}
\newcommand*{\di}{\mathrm{d}}

\renewcommand*{\eqref}[1]{Eq.~(\ref{#1})}
\newcommand*{\Eqref}[1]{Equation~(\ref{#1})}
\newcommand*{\figref}[1]{Fig.~\ref{#1}}

\newcommand*{\secref}[1]{section~\ref{#1}}
\newcommand*{\Secref}[1]{Section~\ref{#1}}
\newcommand*{\appref}[1]{appendix~\ref{#1}}

\newcommand*{\Tabref}[1]{Table~\ref{#1}}

\begin{document}
\title{Quantum Sampling Algorithms, Phase Transitions, and Computational Complexity}
\date{\today}

\author{Dominik S.~Wild}
\affiliation{Max Planck Institute of Quantum Optics, Hans-Kopfermann-Stra{\ss}e 1, D-85748 Garching, Germany}

\author{Dries Sels}
\affiliation{Center for Computational Quantum Physics, Flatiron Institute, New York, New York 10010, USA}
\affiliation{Department of Physics, New York University, New York, New York 10003, USA}

\author{Hannes Pichler}
\affiliation{Institute for Theoretical Physics, University of Innsbruck, Innsbruck A-6020, Austria}
\affiliation{Institute for Quantum Optics and Quantum Information, Austrian Academy of Sciences, Innsbruck A-6020, Austria}

\author{Cristian Zanoci}
\affiliation{Department of Physics, Massachusetts Institute of Technology, Cambridge, Massachusetts 02139, USA}

\author{Mikhail D.~Lukin}
\affiliation{Department of Physics, Harvard University, Cambridge, Massachusetts 02138, USA}

\begin{abstract}
  Drawing independent samples from a probability distribution is an important computational problem with applications in Monte Carlo algorithms, machine learning, and statistical physics. The problem can in principle be solved on a quantum computer by preparing a quantum state that encodes the entire probability distribution followed by a projective measurement. We investigate the complexity of adiabatically preparing such quantum states for the Gibbs distributions of various classical models including the Ising chain, hard-sphere models on different graphs, and a model encoding the unstructured search problem. By constructing a parent Hamiltonian, whose ground state is the desired quantum state, we relate the asymptotic scaling of the state preparation time to the nature of transitions between distinct quantum phases. These insights enable us to identify adiabatic paths that achieve a quantum speedup over classical Markov chain algorithms. In addition, we show that parent Hamiltonians for the problem of sampling from independent sets on certain graphs can be naturally realized with neutral atoms interacting via highly excited Rydberg states.
\end{abstract}

\maketitle

\section{Introduction}

A sampling problem is the task of drawing samples from an implicitly defined probability distribution, which may, for example, be the Gibbs distribution of a classical system at a fixed temperature. This particular class of sampling problems encompasses many practically significant and challenging problems with applications in areas as varied as statistical physics~\cite{Landau2013}, optimization~\cite{Kirkpatrick1983}, and machine learning~\cite{Bishop2006}. Moreover, sampling problems involving Gibbs distributions represent a natural setting to explore connections between phase transitions and computational complexity. It has been shown that Markov chain algorithms with local updates can be used to efficiently sample, in a time polynomial in the system size, from the Gibbs distribution of lattice models if and only if spatial correlations are short ranged~\cite{Martinelli1999,Dyer2004}. This implies that it is easy to sample from many Gibbs distributions at high temperature, while no efficient, general-purpose algorithm is known below the ordering transition temperature of the underlying physical system.

The development of programmable quantum systems raises the question of what connections exist between quantum complexity and phase transitions. In analogy to the statement about classical Gibbs distributions, it has been shown that a quantum computer can efficiently generate samples from quantum Gibbs distributions with short-ranged correlations~\cite{Brandao2018,Kastoryano2016}. Quantum mechanics also offers an alternative motivation to consider sampling problems. Any quantum state $\ket{\psi}$ together with an orthonormal basis $\{\ket{s}\}$ encodes a sampling problem: According to the Born rule, a projective measurement in the $\{\ket{s}\}$ basis yields the outcome $s$ with probability $p(s) = |\inner{s}{\psi}|^2$. By defining $p(s)$ in terms a quantum gate sequence~\cite{Bouland2018} or an optical network~\cite{Aaronson2011}, it is possible to construct sampling problems that can be efficiently solved on a quantum computer but which are hard to solve classically. These efforts have led to impressive experimental demonstrations~\cite{Arute2019,Zhong2020}, showing that current quantum devices may outperform classical devices for specifically tailored sampling problems.

In our recent work, reference~\cite{PRL}, we introduced a family of quantum algorithms that provide unbiased samples by preparing a state that encodes the entire Gibbs distribution. We showed that this approach leads to a speedup over a classical Markov chain algorithm for several examples. In this Article, we explore in detail a novel connection between classical sampling problems and quantum phase transitions that arises in this context. Concretely, we construct a so-called parent Hamiltonian $H_q$, which has the defining property that the state $\ket{\psi}$ encoding the probability distribution of interest is its ground state. The construction starts from a Markov chain that converges to the desired probability distribution. The gap between the ground state and the first excited state of the parent Hamiltonian can be related to the mixing time of the Markov chain, thereby establishing a connection between quantum phases and the classical complexity of the sampling problem. We also investigate the quantum complexity of the problem by exploring the time required to adiabatically prepare the state $\ket{\psi}$. We focus on sampling from classical Gibbs distributions at inverse temperature $\beta$, which naturally define a continuous family of sampling problems as well as a continuous family of states $\ket{\psi(\beta)}$ and Hamiltonians $H_q(\beta)$. Analyzing in detail several examples, we find for each that adiabatic state preparation along the one-parameter family $H_q(\beta)$, which may be viewed as a quantum analogue of simulated annealing~\cite{Somma2007}, exhibits the same time complexity as sampling by means of the Markov chain used to construct the parent Hamiltonian. By contrast, a quantum speedup can be achieved along paths in a suitably extended parameter space for all of the examples. In each case, we explain the magnitude of the speedup in terms of the distinct quantum phases occupying the extended parameter space and the nature of the phase transitions between them.

Our results complement existing quantum algorithms for sampling problems~\cite{Lidar1997,Terhal2000,Somma2007,Somma2008,Wocjan2008,Bilgin2010,Yung2010,Temme2011,Riera2012,Yung2012,Boixo2015,Ge2016,Swingle2016,Harrow2020} by offering a physical picture of quantum speedup. Moreover, our approach is suitable for implementation on near-term quantum devices. We show in particular that the parent Hamiltonian for sampling from independent sets on certain graphs has a natural implementation using highly excited Rydberg states of neutral atoms that is compatible with recently demonstrated programmable atom arrays~\cite{Ebadi2020,Scholl2020}.

This Article is organized as follows. In \secref{sec:parent}, we show how the parent Hamiltonian $H_q(\beta)$ can be constructed from a Markov chain. In \secref{sec:adiabatic}, we discuss general aspects of adiabatic state preparation. This formalism is applied to the first example, the Ising chain, in \secref{sec:ising}. Other examples, namely sampling from independent sets and the unstructured search problem, follow in \secref{sec:wis} and \secref{sec:search}. \Secref{sec:wis} also includes a scheme to experimentally realize the parent Hamiltonians that arise for sampling from independent sets. We provide a summary and outlook in \secref{sec:summary}.

\section{Parent Hamiltonians\label{sec:parent}}
We consider the problem of sampling from the Gibbs distribution of a classical Hamiltonian $H_c$ at inverse temperature $\beta$. Labeling the microstates of the system by $s$, the probability distribution is given by $p(s) = e^{- \beta H_c(s)}/\mathcal{Z}$, where $\mathcal{Z} = \sum_s e^{- \beta H_c(s)}$ is the partition function. The entire Gibbs distribution can be encoded in the quantum state
\begin{equation}
  \ket{\psi(\beta)} = \frac{1}{\sqrt{\mathcal{Z}}} \sum_s e^{- \beta H_c(s)/2} \ket{s},
  \label{eq:gibbs}
\end{equation}
which we will refer to as a Gibbs state. For concreteness, we focus on systems composed of $n$ classical spins such that $s$ may be represented by a string of $n$ bits. It is worth noting that  the Gibbs state can be represented by a projected pair entangled state (PEPS) with bond dimension $D = 2$ under the additional assumption that the Hamiltonian involves only single- and two-body terms~\cite{Verstraete2006}.

Preparing a quantum system in the Gibbs state allows one to obtain an independent sample from the Gibbs distribution associated with $H_c$ by simply performing a projective measurement in the $\{ \ket{s} \}$ basis. Hence, sampling from a classical Gibbs distribution can be reduced to preparing the corresponding Gibbs state. We can establish a connection between the complexity of state preparation and quantum phases by introducing the notion of parent Hamiltonians. A parent Hamiltonian of a state $\ket{\psi}$ is any Hamiltonian for which $\ket{\psi}$ is a ground state. Starting from another Hamiltonian whose ground state can be easily initialized, the Gibbs state can be efficiently prepared adiabatically if the initial Hamiltonian can be smoothly deformed into the parent Hamiltonian without encountering a small energy gap. By contrast, adiabatic state preparation is slow if the Hamiltonian exhibits a small gap (vanishing in the thermodynamic limit) at any point along the adiabatic path. This occurs in particular when the physical system undergoes a quantum phase transition. The parent Hamiltonian of a Gibbs state is not unique. If a PEPS representation of the state is known, there exists a general method of constructing a parent Hamiltonian~\cite{Verstraete2006,Perez-Garcia2008}. For our purposes, it is more convenient to pursue a different construction, where the parent Hamiltonian is obtained from a Markov chain whose stationary distribution is given by the Gibbs distribution~\cite{Verstraete2006}. 

Markov chains constitute a powerful tool to solve sampling problems. To sample from the Gibbs distribution $p(s) = e^{- \beta H_c}/\mathcal{Z}$, we introduce the generator matrix $M$, whose matrix elements $M(s, s')$ specify the transition probability from microstate $s$ to $s'$. A probability distribution $q_t(s)$ at time $t$ evolves into $q_{t+1}(s) = \sum_{s'} q_t(s') M(s', s)$ after one time step of the Markov chain. We impose detailed balance on the Markov chain, which can be expressed as $e^{- \beta H_c(s')} M(s', s) = e^{- \beta H_c(s)} M(s, s')$. Detailed balance ensures that $p(s)$ is a stationary distribution of the Markov chain, that is, $p(s) = \sum_{s'} p(s') M(s', s)$. If, in addition, the Markov chain is fully mixing and aperiodic, the Perron--Frobenius theorem guarantees that $p(s)$ is the unique stationary distribution~\cite{Seneta1981}. Hence, running the Markov chain for a sufficient number of steps yields a sample from the desired Gibbs distribution. The relevant number of steps is known as the mixing time, which can be related to the spectral properties of the generator matrix. Since $M$ is a stochastic matrix, its largest eigenvalue is $\lambda_1 = 1$ with $p(s)$ being the corresponding left eigenvector. Under the assumptions that render the stationary distribution unique, the largest eigenvalue is nondegenerate and the eigenvalue with the second largest magnitude satisfies $|\lambda_2| < 1$. It can be shown that the mixing time satisfies the lower bound $\tau_m \geq |\lambda_2|/(1 - |\lambda_2|)$~\cite{Aldous1982,Levin2009}. This inequality shows that a Markov chain mixes slowly if the spectral gap between the largest and second largest eigenvalue is small.

The generator $M$ can be turned into a parent Hamiltonian of the Gibbs state in \eqref{eq:gibbs} by a similarity transformation following reference~\cite{Verstraete2006} and related earlier work~\cite{Aharonov2003,Henley2004,Castelnovo2005}. As a consequence of detailed balance,
\begin{equation}
  H_q(\beta) = n \left( \mathbb{I} - e^{-\beta H_c / 2} M e^{\beta H_c/2} \right)
  \label{eq:parent}
\end{equation}
is a real, symmetric matrix and thus represents a valid quantum Hamiltonian. The Gibbs state $\ket{\psi(\beta)}$ is an eigenstate of $H_q(\beta)$ with eigenenergy zero, which follows from the fact that $M$ is a right stochastic matrix. We can further show that $\ket{\psi(\beta)}$ is the unique ground state of $H_q(\beta)$ by noting that for every eigenvalue $n (1 - \lambda)$ of $H_q(\beta)$, there exists a corresponding eigenvalue $\lambda$ of $M$. Since the largest eigenvalue $\lambda_1 = 1$ of $M$ is nondegenerate, the ground state of $H_q(\beta)$ is also nondegenerate and has eigenenergy zero. The factor of $n$ in \eqref{eq:parent} ensures that the spectrum of the parent Hamiltonian is extensive. The rescaled spectrum of the parent Hamiltonian as compared to the generator matrix $M$ of the Markov chain reflects the natural parallelization in a quantum system, where each qubit represents a physical resource. For a fair comparison between the mixing time of the Markov chain and the adiabatic state preparation time studied below, we divide the mixing time by $n$, defining $t_m = \tau_m / n$. It then follows from the correspondence of the spectra of $M$ and $H_q(\beta)$ that $t_m \geq 1/\Delta(\beta) - 1/n$, where $\Delta(\beta)$ is the gap between the ground state and first excited state of the parent Hamiltonian. 

The family of Hamiltonians $H_q(\beta)$ can be viewed as generalized Rokhsar--Kivelson Hamiltonians, which are stoquastic and frustration free~\cite{Rokhsar1988,Henley2004,Castelnovo2005,Bravyi2010}. The quantum phases of $H_q(\beta)$ and the phase transitions separating them not only allow us to understand the performance of the Markov chain in physical terms, but they also explain why adiabatic state preparation along certain paths in parameter space offer a quantum speedup over the Markov chain. The achievable speedup depends on the nature of the quantum phase transitions encountered along the adiabatic path and its origin can be traced to quantum phenomena such as ballistic transport and tunneling.

\section{Adiabatic state preparation\label{sec:adiabatic}}
While there exist general approaches to running Markov chains on a quantum computer with a guaranteed quantum speedup~\cite{Somma2008,Wocjan2008,Boixo2015,Ge2016}, these methods require deep circuits that are beyond the reach of current quantum devices. For this reason, we explore the prospect of preparing Gibbs states adiabatically. This alternative avenue has the potential of being realized on near-term devices if the adiabatic evolution can be naturally implemented with little overhead and low noise. Since the adiabatic state preparation time strongly depends on the local rate of change of the Hamiltonian, we now describe a heuristic scheme used to determine the rate of change along a fixed path in parameter space.

The scheme is based on the adiabaticity condition, which states that the population of the instantaneous eigenstates $\ket{n}$ due to nonadiabatic transitions from the eigenstate $\ket{m}$ can be approximately bounded by
\begin{equation}
  p_{m \to n} \lesssim \frac{1}{(E_n - E_m)^2} \left| \bra{n}{\frac{\di}{\di t}}\ket{m} \right|^2,
  \label{eq:transition}
\end{equation}
where $E_m$ and $E_n$ are the respective instantaneous eigenenergies~\cite{Messiah2014,Rezakhani2009}. We require that the transition probability from the ground state into any excited state be small along the adiabatic path. Denoting the ground state by $\ket{0}$ and setting the ground state energy $E_0=0$, we require the sum of the right-hand side of \eqref{eq:transition} over all excited states to satisfy
\begin{equation}
  \sum_{n >0} \frac{1}{E_n^2} \left| \bra{n}{\frac{\di}{\di t}}\ket{0} \right|^2 = \varepsilon^2
  \label{eq:sum}
\end{equation} 
for some small constant $\varepsilon$. For a fixed path in parameter space, \eqref{eq:sum} determines the rate of change of the Hamiltonian parameters.

It is convenient to introduce a dimensionless parameter $s$, which increases monotonically from $s=0$ at the beginning of the path to $s=1$ at the final point. The Hamiltonian is parametrized by a set of parameters $\{ \lambda_\mu\}$. An adiabatic path is fixed by the functional dependence $\lambda_\mu(s)$, while the rate of change of the parameters is captured by the differential equation
\begin{equation}
  \frac{\di t}{\di s} = \frac{1}{\varepsilon} \sqrt{\sum_{\mu, \nu} g_{\mu \nu}(s) \frac{\di \lambda_\mu}{\di s} \frac{\di \lambda_\nu}{\di s}},
  \label{eq:rate}
\end{equation}
which follows immediately from \eqref{eq:sum}, where
\begin{equation}
  g_{\mu \nu} = \sum_{n > 0} \frac{\inner{\partial_\mu 0}{n} \inner{n}{\partial_\nu 0}}{E_n^2}
  \label{eq:metric}
\end{equation}
and $\ket{\partial_\mu 0}$  is a shorthand for $\partial \ket{0} / \partial \lambda_\mu$. The total evolution time can be obtained by integrating \eqref{eq:rate}, yielding
\begin{equation}
  t_\mathrm{tot} = \frac{l}{\varepsilon},
  \label{eq:ttot}
\end{equation}
where
\begin{equation}
  l = \int_0^1 \di s \sqrt{\sum_{\mu, \nu}g_{\mu\nu}(s) \frac{\di \lambda_\mu}{\di s} \frac{\di \lambda_\nu}{\di s}}.
  \label{eq:l}
\end{equation}

The above adiabatic schedule exhibits a number of important features. \Eqref{eq:metric} accounts not only for the gap between the ground state and the first excited state, but also the gap to all other excited states. In this way, we ensure that the adiabatic condition can be satisfied at second-order quantum phase transitions, where an extensive number of states is energetically close to the ground state. In addition, \eqref{eq:metric} takes into account the matrix elements that induce nonadiabatic transitions to appropriately weigh the significance of each excited state. The total evolution time $t_\mathrm{tot}$ can be tuned by adjusting the constant $\varepsilon$. The constant of proportionality $l$ only depends on the path and not its parametrization. For this reason, we refer to $l$ as the adiabatic path length and to $g_{\mu \nu}$ as the adiabatic metric.

To analyze the performance of adiabatic state preparation, we introduce the fidelity
\begin{equation}
  \mathcal{F} = | \inner{\psi(\beta)}{\phi(t_\mathrm{tot})}|^2
\end{equation}
where $\ket{\psi(\beta)}$ is the desired Gibbs state and $\ket{\phi(t_\mathrm{tot})}$ is the final state after evolving under the time-dependent Hamiltonian for a given adiabatic path with the rate of change of the parameters as described above. The fidelity bounds the total variation distance $d = || p - q||$ between the desired Gibbs distribution and the prepared distribution $q(s) = | \inner{s}{\phi(t_\mathrm{tot})}|^2$ by $d \leq \sqrt{1 - F}$~\cite{Nielsen2010}. The adiabatic theorem guarantees that $\mathcal{F} \to 1$ in the limit $t_\mathrm{tot} \to \infty$. Here, we are interested in the shortest evolution time $t_a$ such that the fidelity exceeds some threshold, taken to be $1 - 10^{-3}$ throughout. We will see below that $t_a$ corresponds to a small value of $\varepsilon$ that is approximately independent of the system size.  Hence, the adiabatic path length can serve as a proxy for the total time required to prepare a Gibbs state with high fidelity. One could thus in principle identify good adiabatic paths by minimizing the adiabatic path length, that is, by finding geodesics under the metric $g_{\mu \nu}$. However, for the present purposes, it suffices to compare a small number of paths that uncover distinct asymptotic dependencies of the computation time on the system size.

\section{Ising chain\label{sec:ising}}
\subsection{Parent Hamiltonian\label{sec:ising_parent} and quantum phases}
We now illustrate the formalism outlined above with several concrete examples, starting with the ferromagnetic Ising model in one dimension. To construct the parent Hamiltonian, we choose  Glauber dynamics as the Markov chain, which updates the configurations according to the following prescription~\cite{Glauber1963}: Pick a spin at random and draw its new orientation from the Gibbs distribution with all other spins fixed. Starting from configuration $s$, the probability of flipping spin $i$ in the Ising chain is thus given by
\begin{equation}
  p_i = \frac{1}{2 n \cosh [ \beta (s_{i-1} + s_{i+1})]} e^{- \beta s_i (s_{i-1} + s_{i+1})}.
\end{equation}
By promoting the values of the spins $s_i$ to operators $\sigma_i^z$, we can concisely write the generator of the Markov chain as
\begin{equation}
  M = \sum_i p_i \sigma_i^x + \left(\mathbb{I} - \sum_i p_i \right) .
\end{equation}
\Eqref{eq:parent} then gives
\begin{align}
  H_q(\beta) = \frac{n}{2} \mathbb{I} - \sum_i \left[ h(\beta) \sigma_i^x \right. &+ J_1(\beta) \sigma_{i}^z \sigma_{i+1}^z \nonumber\\
  &- \left. J_2(\beta) \sigma_{i-1}^z \sigma_i^x \sigma_{i+1}^z \right],
  \label{eq:parent_ising}
\end{align}
where $4 h(\beta) = 1 + 1/\cosh(2 \beta)$, $2 J_1 = \tanh(2 \beta)$, $4 J_2(\beta) = 1 - 1/\cosh(2 \beta)$ (see~\cite{Felderhof1971,Siggia1977} for early derivations of this result).

\begin{figure}[t]
  \centering
  \includegraphics{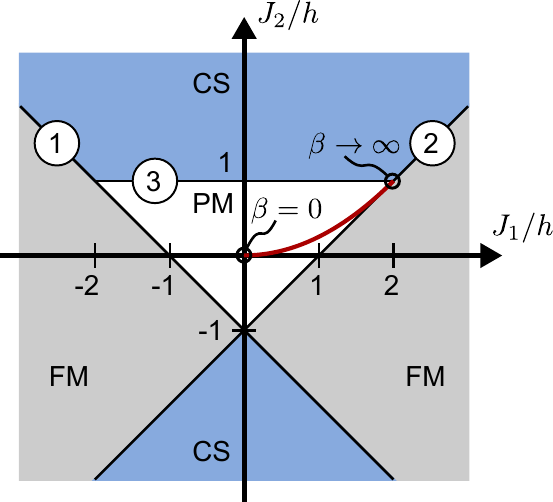}
  \caption{Phase diagram of the Hamiltonian in~\eqref{eq:parent_ising} with arbitrary coefficients ($h > 0$). The colored regions distinguish the paramagnetic (PM), ferromagnetic (FM), and cluster-state-like (CS) phases. The circled numbers indicate which of the cases described in the main text is responsible for closing the gap at the phase transition. The red curve indicates the one-parameter family $H_q(\beta)$, i.e., the parent Hamiltonian of Gibbs state $\ket{\psi(\beta)}$ of the Ising chain.}
  \label{fig:fig1}
\end{figure}

While \eqref{eq:parent_ising} defines a one-parameter family of Hamiltonians dependent on $\beta$, it is natural to consider the quantum phase diagram of $H_q$ in an extended parameter space with arbitrary values of $h$, $J_1$, and $J_2$.  The Hamiltonian in \eqref{eq:parent_ising} is exactly solvable using a Jordan--Wigner transformation to map it onto free fermions (see \appref{app:fermion} for details)~\cite{Skrovseth2009,Niu2012}. One finds that the energy of an excitation with momentum $k$ is proportional to
\begin{align}
  E_k &= \left\{ [h + J_1 \cos (k) + J_2 \cos(2 k)]^2 \right. \nonumber\\
    & \hspace{2cm}+ \left. [J_1 \sin(k) + J_2 \sin(2 k)]^2 \right\}^{1/2}.
    \label{eq:dispersion}
\end{align}
At a quantum phase transition, the energy gap between the ground state and the first excited state vanishes, which implies that there must exist at least one mode with zero energy, i.e., $E_k = 0$ for some $k$. This is indeed the case at $k = 0$ when $h + J_1 + J_2 = 0$ (line 1 in \figref{fig:fig1}), $k = \pi$ when $h - J_1 + J_2 = 0$ (line 2 in \figref{fig:fig1}), or at $k = \pi - \cos^{-1} \frac{J_1}{2 J_2}$ when $h - J_2 = 0$ and $|J_1| < 2 |h|$ (line 3 in \figref{fig:fig1}). To identify the distinct phases, we consider line cuts through the parameter space. Along $J_2 = 0$, the Hamiltonian in \eqref{eq:parent_ising} reduces to the transverse-field Ising model, which exhibits a well-know phase transition from a paramagnetic phase to a ferromagnetic phase at $|J_1| = |h|$. Similarly, the line $J_1 = 0$ has been studied previously, allowing us to identify the remaining third phase as a cluster-state-like phase separated from the paramagnetic phase by a symmetry-protected topological phase transition~\cite{Son2011,Verresen2017}.

All three quantum phases are gapped and the gap only vanishes at the phase transitions. The dispersion relation given by \eqref{eq:dispersion} is linear at low energies for all phase transitions with the exception of the tricritical points at $J_1/h = \pm 2$, $J_2/h = 1$. At the tricritical points, the low-lying excitations from two distinct phase transitions merge into a single gapless mode with a quadratic dispersion relation. This behavior is captured by the dynamical critical exponent $z$~\cite{Sachdev2009}, which takes the value $z = 2$ at the tricritical points and $z = 1$ everywhere else. For a finite-sized system of $n$ spins, the gap closes as $\sim n^{-z}$ since the spacing of momenta is inversely proportional to $n$. We note that at the transition between the paramagnetic and cluster-state-like phases, the gap displays an oscillatory behavior as a function of system size and may even vanish exactly. Nevertheless, the envelope follows the expected $\sim n^{-1}$ scaling.  

The above insights into the phase diagram and the nature of the phase transitions have immediate implications on both the mixing time of Markov chain and the adiabatic state preparation time. The parent Hamiltonian $H_q(\beta)$ (red curve in \figref{fig:fig1}) occupies the paramagnetic phase and is therefore gapped for all finite $\beta$. Hence, the Markov chain is expected to mix rapidly in this regime as confirmed by the rigorous bound $t_m \lesssim \log n$~\cite{Levin2009}. Similarly, adiabatic state preparation starting from any other state in the paramagnetic regime is efficient~\cite{Ge2016}. By contrast, the parent Hamiltonian of the zero temperature ($\beta \to \infty$) Gibbs state resides at a tricritical point in the phase diagram, whose dynamical critical exponent bounds the mixing time by $t_m \gtrsim n^2$. This scaling can be understood as a consequence of the diffusive propagation of domain walls in the classical Markov chain dynamics. For the Markov chain to converge to the perfect ferromagnetic states with all spins aligned, it is necessary to remove all domain boundaries. Since the largest domains can be on the order of the system size, removing them will in general take on the order of $n^2$ steps in the random walk of the domain wall. The possibility of ballistic propagation in a coherent quantum system, as evidenced by the dynamical critical exponent $z = 1$ away from the tricritical point, suggests that adiabatic state preparation could achieve a quadratic speedup over the classical Markov chain for sampling from the Gibbs distribution of the Ising chain at zero temperature. We show in the next section that this simple argument indeed captures the relevant physics.

\subsection{Adiabatic state preparation time}
\begin{figure}[t]
  \centering
  \includegraphics{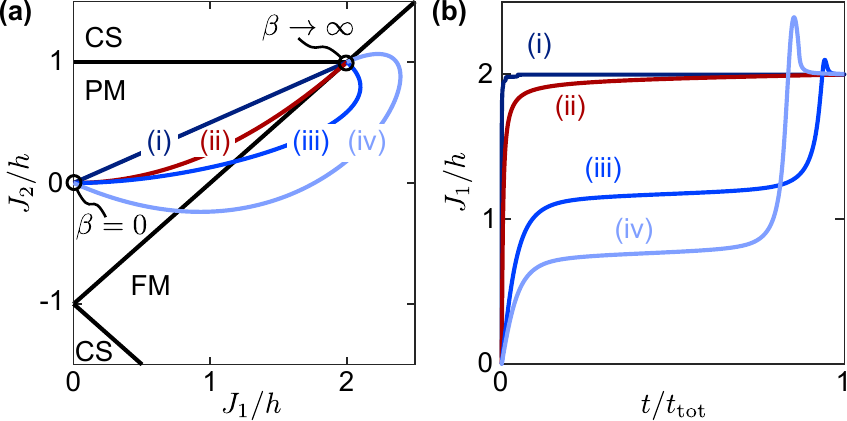}
  \caption{Adiabatic paths to prepare the zero-temperature Gibbs state of the Ising chain. (a) The four paths (i)--(iv) all start at $H_q(0)$ and end at $H_q(\beta)$ with $\beta \to \infty$. Path (ii) corresponds to the one-parameter family of parent Hamiltonians $H_q(\beta)$. An explicit parametrization of the paths is provided in \appref{app:ising_paths}. (b) Time dependence of $J_1/h$ along the four paths in (a) for a chain of $n = 100$ spins according to the scheme outlined in \secref{sec:adiabatic}.}
  \label{fig:fig2}
\end{figure}

To adiabatically prepare the Gibbs state at any temperature, we start from the ground state of the Hamiltonian \eqref{eq:parent_ising} with $J_1/h = J_2/h = 0$, corresponding to the parent Hamiltonian at infinite temperature ($\beta = 0$). The ground state is a product state of each spin aligned along the $x$ direction, which, we assume, can be easily prepared. A measurement of this state in the computational basis yields any configuration with equal probability as required for a Gibbs distribution at infinite temperature. As argued above, adiabatic state preparation of the Gibbs sate $\ket{\psi(\beta)}$ proceeds efficiently for any finite value of $\beta$. We therefore focus on the more challenging problem of sampling from the Gibbs distribution of the Ising chain at zero temperature, investigating the four adiabatic paths in \figref{fig:fig2}(a).

Before we proceed with the detailed analysis, we comment on an issue related to the breaking of the $\mathbb{Z}_2$ symmetry by the ferromagnetic ground state. The Markov chain is in fact nonergodic at zero temperature and the gap of the generator matrix vanishes exactly. The reason for this is that both ferromagnetic configurations (all spins up or down) are stationary states of the Markov chain. It is nevertheless possible to sample from the stationary distribution by restarting the Markov chain in a random configuration. Rather than specifying a mixing time, which is ill defined, we should characterize the wait time that is necessary between restarts. Fortunately, the quantum formalism avoids this complication entirely as the system remains in the even parity subspace (see \appref{app:fermion}) at all times. The ground state of the even parity subspace remains unique, while the nonergodicity of the Markov chain is reflected in the degeneracy between the even and odd subspaces. The relevant wait time for the Markov chain can be estimated from the gap of the Hamiltonian in the even subspace. Since the distinction between the wait and the mixing time is irrelevant from the perspective of computational complexity, we will use the latter term to encompass both and we will refer to them by the same symbol $t_m$ in what follows.

\begin{figure}[t]
  \centering
  \includegraphics{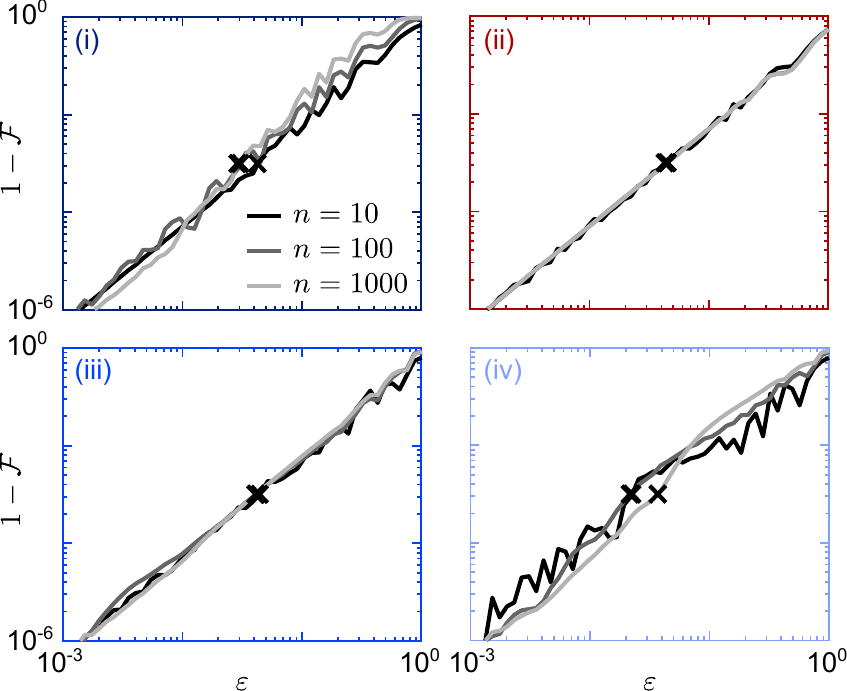}
  \caption{Infidelity $1 - \mathcal{F}$ as a function of $\varepsilon$ for the four paths in \figref{fig:fig2}(a). Three chain lengths $n = 10$, $100$, and $1000$ are shown, see the legend in the top-left panel. The black crosses indicate where the infidelity equals $10^{-3}$.}
  \label{fig:fig3}
\end{figure}

\begin{figure}[t]
  \centering
  \includegraphics[width=\columnwidth]{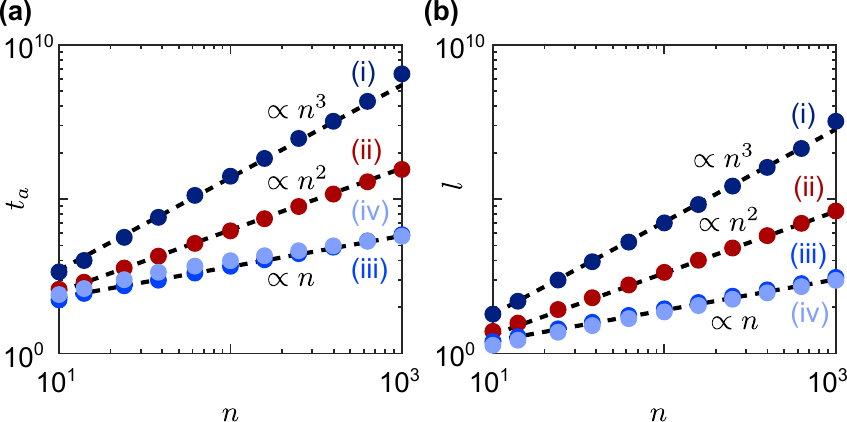}
  \caption{Adiabatic state preparation of the zero-temperature Gibbs state of the Ising chain as a function of the chain length $n$. (a) Adiabatic state preparation time $t_a$ as a function of the chain length $n$. The dots show the numerical result for the four paths in \figref{fig:fig2}(a). (b) The adiabatic path length $l$ agrees with the adiabatic state preparation time up to a constant prefactor. In both panels, the dashed lines are guides to the eye showing the power laws $\propto n$, $n^2$, $n^3$.}
  \label{fig:fig4}
\end{figure}

To compute the adiabatic state preparation time for each path in \figref{fig:fig2}(a), we numerically integrate the time-dependent Schrödinger equation, where the rate of change of the parameters is chosen following \secref{sec:adiabatic} (see \appref{app:tdse} for details). The resulting dependence of $J_1/h$ on $t/t_\mathrm{tot}$ is plotted in \figref{fig:fig2}(b) in the case of $n = 100$ spins. We obtain the fidelity $\mathcal{F}$ as a function of $\varepsilon$, which determines the total evolution time $t_\mathrm{tot}$ according to \eqref{eq:ttot}, for different system sizes ranging from $n = 10$ to $n = 1000$. As can be seen from \figref{fig:fig3}, the fidelity depends only weakly on the system size and exhibits a universal dependence on $\varepsilon$ with $1 - \mathcal{F}$ approximately proportional to $\varepsilon^2$. This can be understood from the fact that the first-order diabatic correction to the wavefunction is inversely proportional to the total time $t_{\mathrm{tot}} = l/\varepsilon$~\cite{Messiah2014}. From these results, we compute the adiabatic state preparation time $t_a$ as the earliest time $t_\mathrm{tot}$ at which $1 - \mathcal{F}$ is less than $10^{-3}$. The result is shown in \figref{fig:fig4}(a). 

The adiabatic state preparation time of the four paths follows distinct dependencies on system size. Path (ii), which corresponds to the one-parameter family $H_q(\beta)$, exhibits the scaling $t_a \sim n^2$. This matches the scaling of the lower bound on the sampling time of the Markov chain at zero temperature. Paths (iii) and (iv), which cross into the ferromagnetic phase, achieve a quadratically faster scaling, $t_a \sim n$. This scaling is in agreement with our earlier prediction that crossing the phase transition away from the tricritical point facilitates ballistic propagation of domain walls. Finally, path (i) is slower with $t_a \sim n^3$. We will see below that this slowdown can be attributed to the fact that the gap at the transition between the paramagnetic phase and the cluster-state-like phase may close exactly even for finite-sized system.

We may gain analytic insight into the above scalings by considering the adiabatic path length $l$. Since the fidelity is largely independent of $n$ for fixed $\varepsilon$, the adiabatic path length serves as a reliable proxy for the adiabatic state preparation time $t_a$. Indeed, \figref{fig:fig4}(b) shows that $l$ and $t_a$ exhibit roughly the same dependence on system size up to a constant prefactor. The adiabatic path length is dominated by the singular behavior close to the tricritical point. We show in \appref{app:l} that in the thermodynamic limit, the adiabatic metric diverges at the tricritical point as
\begin{align}
  \label{eq:G}
  G &= \sum_{\mu, \nu} g_{\mu \nu} \frac{\di \lambda_\mu}{\di \eta} \frac{\di \lambda_\nu}{\di \eta} \\
  &\approx \frac{3 n}{2048} \times
  \begin{cases}
    \frac{1}{32} \frac{\sin^2 \alpha}{(\cos \alpha - \sin \alpha)^{5/2}} \eta^{-5/2} + O(\eta^{-3/2})\\
    \frac{1}{|\sin^3 \alpha|} \eta^{-5} + O(\eta^{-4})
  \end{cases}. \nonumber
\end{align}
Here $\eta$ and $\alpha$ specify the distance and direction relative to the tricritical point: $J_1 = 2 + \eta \cos \alpha$, $J_2 = 1 + \eta \sin \alpha$, having set $h = 1$. The first case in \eqref{eq:G} corresponds to the ferromagnetic phase ($-3 \pi/4 < \alpha < \pi / 4$), while the second case applies to the paramagnetic and cluster-state-like phases ($\pi / 4 < \alpha < 5 \pi / 4$). In both cases, the adiabatic metric diverges as a power law, $G \sim n \eta^{-\rho}$, with $\rho = 5/2$ in the ferromagnetic phase and $\rho = 5$ otherwise.

For finite-sized systems, one can show that exactly at the critical point, $G \sim n^\sigma$, where $\sigma = 6$ in any direction not parallel to the $J_1$ axis. Based on finite-sized scaling arguments, we expect that the metric follows the expression for the infinite system as we approach the tricritical point until it saturates to the final value $G \sim n^\sigma$. We are thus led to define a critical region $\eta < \eta_c$ determined by $n \eta_c^{-\rho} \sim n^{\sigma}$, in which the metric is approximately constant.  These arguments imply that the path length should scale according to
\begin{equation}
  l \sim n^{(2 - 2 \sigma + \rho \sigma)/(2 \rho)} \sim
  \begin{cases}
    n, & -\frac{3 \pi}{4} < \alpha < \frac{\pi}{4}\\
    n^2, & \phantom{-} \frac{\pi}{4} < \alpha < \frac{5 \pi}{4}
  \end{cases}.
\end{equation}
The above prediction agrees well with the numerical results for paths (ii)--(iv) in \figref{fig:fig4}, however, it fails for path (i), for which a cubic scaling is observed. In the latter case, the scaling hypothesis breaks down since the dispersion minimum is neither at $k = 0$ or $k = \pi$ as the tricritical point is approached from the ferromagnetic phase, thereby giving rise to an incommensurate length scale that breaks scale invariance. A similar analysis can be performed at the transition between the paramagnetic and the ferromagnetic phases, away from the tricritical point. One finds that the adiabatic path length always scales linearly with the system size, in agreement with the numerical results for paths (iii) and (iv).

\section{Sampling from weighted independent sets\label{sec:wis}}
\subsection{Parent Hamiltonian}
As the next example, we consider the problem of sampling from weighted independent sets. Given an undirected graph with vertices $V$, an independent set is any subset of $V$ in which no pair of vertices is connected by an edge. We assign to each independent set the classical energy
\begin{equation}
  H_c = - \sum_i w_i n_i,
  \label{eq:hc_wis}
\end{equation}
where the sum runs over all vertices and $w_i$ are positive weights. The occupation number $n_i=1$ (occupied) when the vertex $i$ is in the independent set and $n_i = 0$ (unoccupied) otherwise. The task at hand is to sample from the Gibbs distribution associated with $H_c$ for a given inverse temperature $\beta$. Problems of this type encompass a large class of challenging problems that appear in diverse practical settings. For instance, finding the maximum independent set, which may be viewed as sampling at zero temperature with equal weights $w_i = 1$, has applications in computer vision~\cite{Bomze1999}, biochemistry~\cite{Butenko2006}, and social network analysis~\cite{Fortunato2010}. The problem is hard for general graphs even at a finite (but sufficiently low) temperature since approximately solving the maximum independent set problem on general graphs is NP hard~\cite{Garey1979,Sly2010}. The Hamiltonian in \eqref{eq:hc_wis} is also interesting from a physics perspective as a model of a lattice gas of hard spheres~\cite{Gaunt1965,Weigt2001}.

To derive a parent Hamiltonian for the Gibbs state, we construct a Markov chain with the Metropolis--Hastings update rule~\cite{Hastings1970}. A move is accepted with probability $p_\mathrm{accept} = \min (1, e^{-\beta \Delta E})$, where $\Delta E$ is the change in energy. With single site updates, the probability of changing the occupation of vertex $i$ is given by
\begin{equation}
  p_i = P_i e^{- \beta w_i n_i}.
\end{equation}
The projector $P_i = \prod_{j \in \mathcal{N}_i} (1 - n_j)$ projects onto configurations where the nearest neighbors $\mathcal{N}_i$ of vertex $i$ are all unoccupied to ensure that the Markov chain never leaves the independent set subspace. We do not consider the possibility in which the Markov chain is initialized in a state outside this subspace. Following the same steps as in \secref{sec:ising_parent}, we derive the parent Hamiltonian
\begin{align}
  H_q(\beta) &= \sum_i P_i \left[ V_{e,i}(\beta) n_i + V_{g,i}(\beta) (1 - n_i) \right.\nonumber\\
  & \hspace{4.5cm} \left. - \Omega_i(\beta) \sigma_i^x \right],
  \label{eq:parent_wis}
\end{align}
where $V_{e,i}(\beta) =  e^{- \beta w_i }$, $V_{g,i}(\beta) = 1$, and $\Omega_i(\beta) = e^{- \beta w_i/2}$. The Pauli operator $\sigma_i^x$ changes the occupation number of site $i$. For a complete description in terms of spins, we further introduce the Pauli $z$ operators $\sigma_i^z = 1 - 2 n_i$. We remark that constrained quantum models of this type have recently attracted great interest in the context of many-body scars~\cite{Turner2018}. 

\subsection{Implementation with Rydberg atoms}
\begin{figure}[t]
  \centering
  \includegraphics{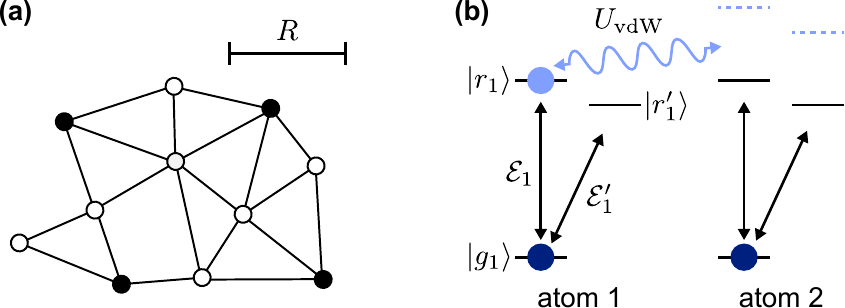}
  \caption{Physical realization of the parent Hamiltonian for sampling from weighted independent sets of unit disk graphs. (a)~An example of an independent set (filled circles) of a unit disk graph. Two vertices are connected if and only if they are separated by a distance less than $R$. (b)~The individual terms are realized by driving optical transitions to highly excited Rydberg states $\ket{r_i}$ and $\ket{r_i'}$, whose strong van der Waals interaction gives rise to Rydberg blockade.}
  \label{fig:fig5}
\end{figure}

Before exploring the quantum phase diagram of \eqref{eq:parent_wis} for particular graphs, we introduce a scheme to realize this Hamiltonian in a system of trapped, neutral atoms interacting via highly excited Rydberg states. The proposal extends a previous method to encode the maximum independent set problem~\cite{Pichler2018a} and it is well suited for programmable atom arrays~\cite{Labuhn2016,Bernien2017,Ebadi2020,Scholl2020}. As a key feature of the scheme, Rydberg blockade allows for direct realization of the projectors $P_i$, which represent $d$-body operators for a vertex of degree $d$, provided the underlying graph is a unit disk graph. This class of graphs consists of vertices embedded in two-dimensional Euclidean space, where two vertices are connected by an edge if and only if they are separated by less than a unit distance $R$ [see \figref{fig:fig5}(a)].

An atom is placed on each vertex of the graph. As illustrated in \figref{fig:fig5}(b), each atom has a ground state $\ket{g_i}$, encoding the unoccupied vertex $i$, and a Rydberg state $\ket{r_i}$ corresponding to the vertex being occupied. The Rydberg blockade ensures that the states of the atoms respect the independent set constraint. More concretely, we implement the first and last term in \eqref{eq:parent_wis} by driving a transition from $\ket{g_i}$ to $\ket{r_i}$. The value of $V_{e,i}$ is set by the detuning of the drive, whereas $\Omega_i$ is proportional to the drive amplitude $\mathcal{E}_i$. The projectors $P_i$ arise naturally from Rydberg blockade: If an atom is excited to the Rydberg state, the strong van der Waals interaction $U_\mathrm{vdW}$ shifts the Rydberg states of all neighboring atoms out of resonance, effectively turning off the drive and thereby enforcing the independent set constraint. The remaining second term in \eqref{eq:parent_wis} can be realized using a similar approach, combining the Rydberg blockade with an AC Stark shift induced by an off-resonant drive from the ground state to an auxiliary Rydberg state $\ket{r_i'}$. The Rydberg interaction contributes additional terms to the Hamiltonian that decay as $1/r^6$ with the distance $r$ between two atoms. We have neglected these terms, noting that a strategy to mitigate their role has been proposed in a related context~\cite{Pichler2018b}. 

\subsection{Chain graph}
We now consider a chain graph of length $n$ with periodic boundary condition and equal weights, $w_i = 1$. The parent Hamiltonian in \eqref{eq:parent_wis} can be written as
\begin{align}
  H_q(\beta) &= \sum_{i=1}^n \left[ (V_e(\beta) - 3 V_g(\beta)) n_i  + V_g(\beta) n_{i-1} n_{i+1} \right.\nonumber\\
  & \hspace{1.2cm} \left.- \Omega(\beta) (1 - n_{i-1}) \sigma_i^x (1 - n_{i+1}) \right],
  \label{eq:pxp}
\end{align}
where we dropped the subscripts $i$ since the coefficients are translationally invariant. We further replaced $P_i n_i$ by $n_i$, which has no effect for independent sets. \Eqref{eq:pxp} has been previously discussed as a model of strongly interacting bosons~\cite{Fendley2004,Lesanovsky2012}. Since the chain graph is a unit disk graph, \eqref{eq:pxp} can in principle be experimentally realized using the scheme described above. An alternative, approximate realization of the Hamiltonian has already been demonstrated by directly taking advantage of the next-to-nearest-neighbor interactions~\cite{Bernien2017}.

Intriguingly, an exact expression for the low-energy states can be obtained along the one-parameter family defined by $\beta$~\cite{Lesanovsky2012}. On the other hand, there is no known exact solution in the extended parameter space spanned by $(\Omega/V_g, V_e/V_g)$, where we assume $V_g > 0$ throughout. We instead obtain an approximate phase diagram by numerically diagonalizing the Hamiltonian for finite chains. The complexity of the problem is reduced as we only need to consider the subspace of independent sets. Given a chain of $n$ vertices with periodic boundary condition, the dimension of the Hilbert space is equal to $F_{n-1} + F_{n+1}$, where $F_n$ is the $n^\mathrm{th}$ Fibonacci number. We can further restrict the analysis to states that are invariant under translation (zero momentum). Assuming that the initial state satisfies this condition, the state will remain in this subspace at all times since the Hamiltonian is translationally invariant. With $n = 24$ vertices, the dimension of this subspace is $d = 4341$, while for $n = 30$, we have $d = 62075$. 

\begin{figure}[t]
  \centering
  \includegraphics{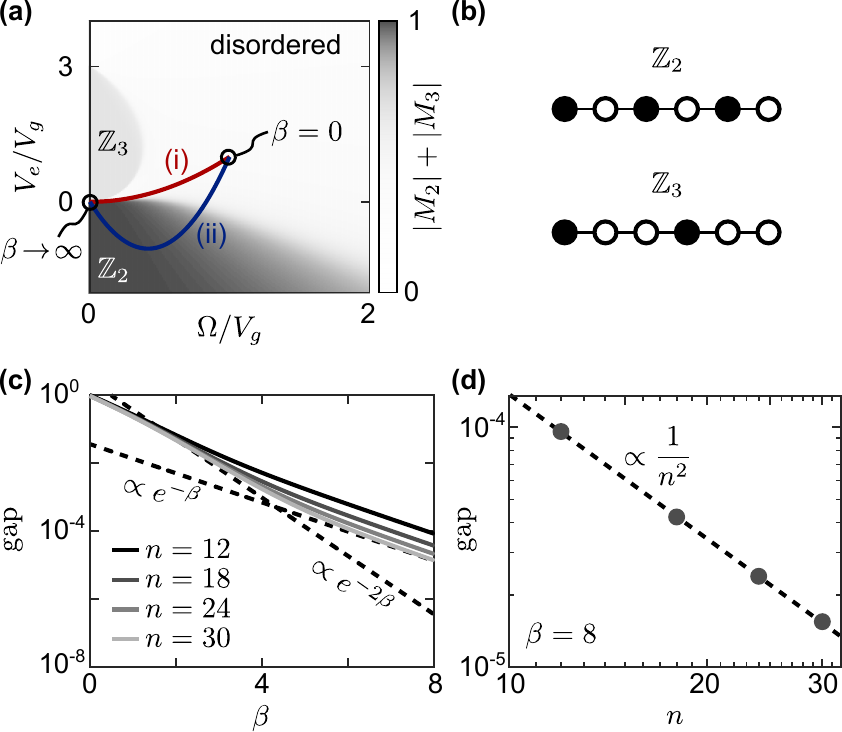}
  \caption{Phase diagram corresponding to sampling from independent sets of a chain graph. (a) Order parameter $|M_2| + |M_3|$ computed for $n=30$ vertices revealing the approximate quantum phase diagram of the Hamiltonian in \eqref{eq:parent_wis} for the chain graph. We study adiabatic state preparation along paths (i) and (ii), an explicit parametrization of which can be found in \appref{app:wis_paths}. (b)~Illustration of the translational symmetry breaking in the $\mathbb{Z}_2$ and $\mathbb{Z}_3$ phases. (c)~Gap of the parent Hamiltonian $H_q(\beta)$ in \eqref{eq:parent_wis} for four different system sizes. The dashed lines show the functions $e^{-2 \beta}$ and $e^{- \beta}$ up to a constant prefactor. (d)~Dependence of the gap on the system size $n$ at $\beta = 8$. The dashed line indicates the power law $1/n^2$.}
  \label{fig:fig6}
\end{figure}

We emphasize that the restriction to the translationally invariant subspace does not exclude spatially ordered states, as they can form a translationally invariant state by equal superpositions of the translated spatial order. It was shown in reference~\cite{Fendley2004} that \eqref{eq:pxp} supports distinct phases with broken translational symmetry, among them a $\mathbb{Z}_2$ and a $\mathbb{Z}_3$ ordered phase [\figref{fig:fig6}(b)], where the spatial order has a period of two or three vertices, respectively. To determine the approximate location of the former two phases, we introduce the order parameter
\begin{equation}
  M_k = \frac{1}{n} \sum_{j = 1}^n e^{2 \pi i j/k} \sigma_j^z
\end{equation}
for integer $k$. By numerically evaluating the expectation value of $|M_2| + |M_3|$ in the ground state of a chain with $n = 30$ sites, we can clearly distinguish the $\mathbb{Z}_2$ and $\mathbb{Z}_3$ phases from the disordered phase [\figref{fig:fig6}(a)]. The phase boundaries are in agreement with those obtained in reference~\cite{Fendley2004} with the caveat that we do not resolve the incommensurate phase that occurs in a thin slice separating the disordered phase from the $\mathbb{Z}_3$ phase.

The one-parameter family $H_q(\beta)$ is indicated by the red curve (i) in \figref{fig:fig6}(a). For large $\beta$, the curve asymptotes to the phase boundary between the disordered phase and the $\mathbb{Z}_2$ phase, which was determined analytically in reference~\cite{Fendley2004}. The spectral gap along the one-parameter family, plotted in \figref{fig:fig6}(c), allows us to bound the mixing time of the Markov chain.  At high temperature, the dependence $e^{- 2 \beta}$ is theoretically expected~\cite{Lesanovsky2012}. Our numerical results are consistent with this prediction, although in the zero-momentum subspace it only holds for large systems. At low temperature, the gap is proportional to $e^{-\beta} / n^2$ as can be seen from \figref{fig:fig6}(c) and (d). Similar to the one-dimensional Ising model, the quadratic dependence on $n$ is a consequence of the diffusive propagation of domain walls in the ordered phase, i.e.~pairs of unoccupied sites that break up the $\mathbb{Z}_2$ ordering. In contrast to the Ising model, these domain walls must overcome an energy barrier to propagate by removing the occupation of an adjacent vertex.  This results in the additional factor $e^{-\beta}$, which implies that the Markov chain is nonergodic at zero temperature even after accounting for the degeneracy that arises from spontaneous symmetry breaking. Unlike for the Ising chain, it is not sufficient to restart the Markov chain to sample from the Gibbs distribution at zero temperature.

While sampling at zero temperature is not possible, we can obtain samples that are close to the zero-temperature distribution by running the chain at a temperature that decreases with system size. The above analysis of the gap indicates that finite size effects, which are due to the correlation length reaching the system size, become relevant when $e^{-2 \beta} \sim e^{- \beta}/n^2$. This suggests that running the Markov chain at inverse temperature $\beta_c = 2 \log n$ will yield a high overlap with the zero-temperature Gibbs distribution. Indeed, we verified numerically that the fidelity of the ground state of the parent Hamiltonian at this temperature relative to the Gibbs state at zero temperature is approximately $90\%$, independent of system size. The constant overlap reflects the fact that the correlation length at $\beta_c$ is a fixed ratio of the system size. It is possible to increase the overlap by adding a constant to $\beta_c$ without changing the scaling behavior discussed in what follows. From the gap of the parent Hamiltonian, we bound the mixing time of the Markov chain at this temperature by $t_m \gtrsim n^4$.

The mixing time is again to be compared to the adiabatic state preparation time. We first observe that the Gibbs state at $\beta = 0$, which is an equal superposition of all independent sets, is not a simple product state. Nevertheless, it can be connected to a product state by a path that lies fully in the disordered phase. For example, the ground state for $\Omega/V_g = 0$, $V_e/V_g > 3$ is a product state of all sites unoccupied. We assume that this state can be readily prepared and thus, by adiabatic evolution along a path that remains in the disordered phase, any Gibbs state corresponding to a nonzero temperature (independent of $n$) can be prepared efficiently. The gapped spectrum of the disordered phase similarly implies that the Markov chain mixes fast for all such states.

\begin{figure}[t]
  \centering
  \includegraphics[width=\columnwidth]{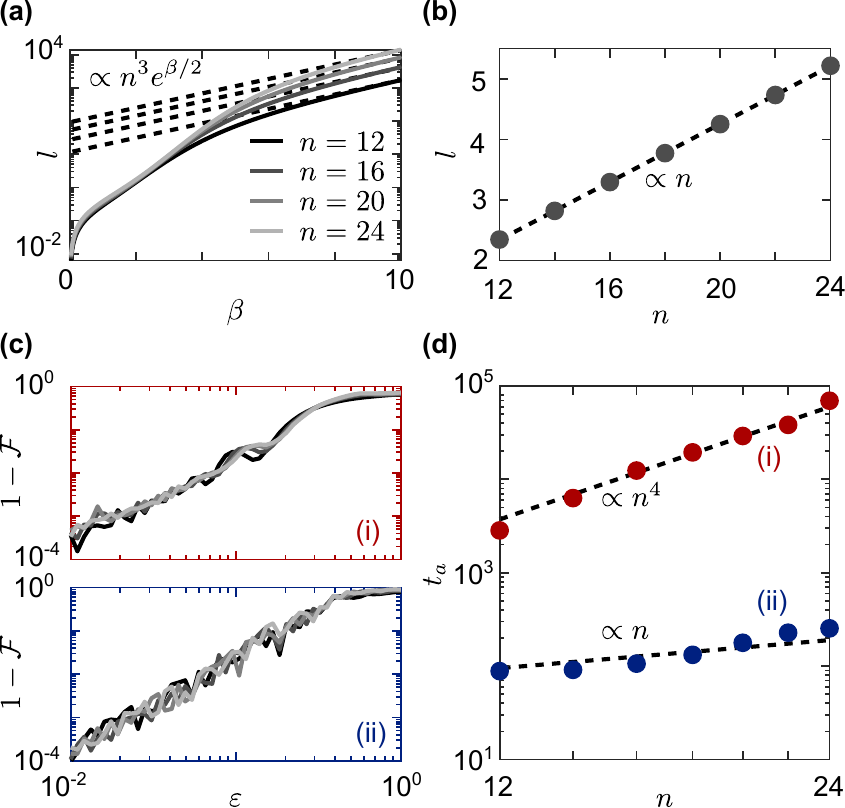}
  \caption{Adiabatic state preparation of the (approximate) zero-temperature Gibbs state for sampling from independent sets of the chain graph. (a) Adiabatic path length along the one-parameter family $H_q(\beta)$, path (i) in \figref{fig:fig6}(a), with the path starting at $0$ and ending at $\beta$. The dashed lines show fits to the value of the function $A e^{\beta/2} n^3$ to the value of $l$ at $\beta = 10$. (b) Adiabatic path length along path (ii) in \figref{fig:fig6}(a), which ends at the origin. The dashed line shows a linear dependence of $l$ on $n$. (c) Infidelity $1 - \mathcal{F}$ as a function of $\varepsilon$ for continuous evolution along the two paths using the rate of change described in \secref{sec:adiabatic}. The differently colored lines correspond to the system sizes specified by the legend in panel (a). (d) Adiabatic state preparation time $t_a$ for the two paths in \figref{fig:fig6}(a). The dashed lines are guides to the eye indicating the power laws $t_a \propto n^4$ and $t_a \propto n$. Note that the target state along path (i) is $\ket{\psi(\beta_c)}$ with $\beta_c = 2 \log n$, while path (ii) aims to reach the exact zero-temperature Gibbs state.}
  \label{fig:fig7}
\end{figure}

Just as for classical sampling using the Markov chain, adiabatically preparing the zero-temperature Gibbs state is challenging. The adiabatic path length $l$ along path (i), i.e. the one-parameter family $H_q(\beta)$, diverges as $n^3 e^{\beta/2}$ as shown in \figref{fig:fig7}(a). It is therefore impossible to reach the zero temperature Gibbs state using the adiabatic schedule described in \secref{sec:adiabatic}. If we follow the same strategy as for the Markov chain and instead prepare the Gibbs state at $\beta_c = 2 \log n$, the relevant adiabatic path length along path (i) depends on the system size as $l \sim n^4$. Intriguingly, there exist other paths, such as path (ii) in \figref{fig:fig6}(a), along which the zero-temperature Gibbs state can be reached from the disordered phase with a finite adiabatic path length. Numerically, we find that the adiabatic path length along path (ii) exhibits the scaling $l \sim n$ [\figref{fig:fig7}(b)]. The path length is dominated by the contribution from the phase transition between the disordered phase and the $\mathbb{Z}_2$ ordered phase. This transition is in the same universality class as the second order phase transition between the paramagnetic and ferromagnetic phases in the transverse-field Ising model~\cite{Fendley2004}. As for the Ising model in \secref{sec:ising}, the linear dependence of the adiabatic path length on the system size is thus a consequence of the dynamical critical exponent $z = 1$ or, more physically, the ballistic propagation of domain wall boundaries at the phase transition.

Assuming that the adiabatic state preparation time is proportional to the adiabatic path length, as was the case for the Ising chain \secref{sec:ising}, we expect $t_a \sim n^4$ along path (i) and $t_a \sim n$ for path (ii). We verify that this assumption holds in \figref{fig:fig7}(c), where we numerically integrated the Schrödinger for chains with length ranging from $n = 12$ to $n = 24$ (see \appref{app:wis_numerical} for details). Hence, the adiabatic state preparation time $t_a$ displayed in \figref{fig:fig7}(d) exhibits the same dependence on system size as the adiabatic path length, $t_a \sim n^4$ for path (i) and $t_a \sim n$ for path (ii). We reiterate that along path (i), we only aim to prepare the Gibbs state $\ket{\psi(\beta_c)}$ with $\beta_c = 2 \log n$, which has a fidelity of approximately $90 \%$ relative to the zero-temperature Gibbs state. No such caveat applies to path (ii), for which the zero-temperature Gibbs is the target state. 

To summarize, adiabatic state preparation along path (ii) achieves a quartic speedup over the classical Markov chain to sample from the Gibbs state at zero temperature. The linear scaling of the quantum algorithm with system size is explained by the ballistic propagation of domain walls, whereas the quartic scaling of the classical algorithm is the result of diffusive propagation combined with a thermal barrier at low temperature. We remark that it is possible to remove the thermal barrier in the Markov chain by introducing an update that simultaneously changes the occupation of two adjacent vertices. With such an update, one can sample from the zero-temperature Gibbs state using the Markov chain in a time $t_m \gtrsim n^2$, limited by diffusion without an energy barrier as in the Ising chain. We do not expect the pair updates to modify the scaling of the adiabatic state preparation time along path (i), which is not limited by diffusion but by the (optimal) ballistic propagation of defects. Thus, the quantum algorithm retains a quadratic speedup when pair updates are included in the Markov chain. It is intriguing that the quantum algorithm achieves the optimal scaling even though the parent Hamiltonian was derived for a suboptimal Markov chain.

\subsection{Star graph}
\begin{figure}[t]
  \centering
  \includegraphics{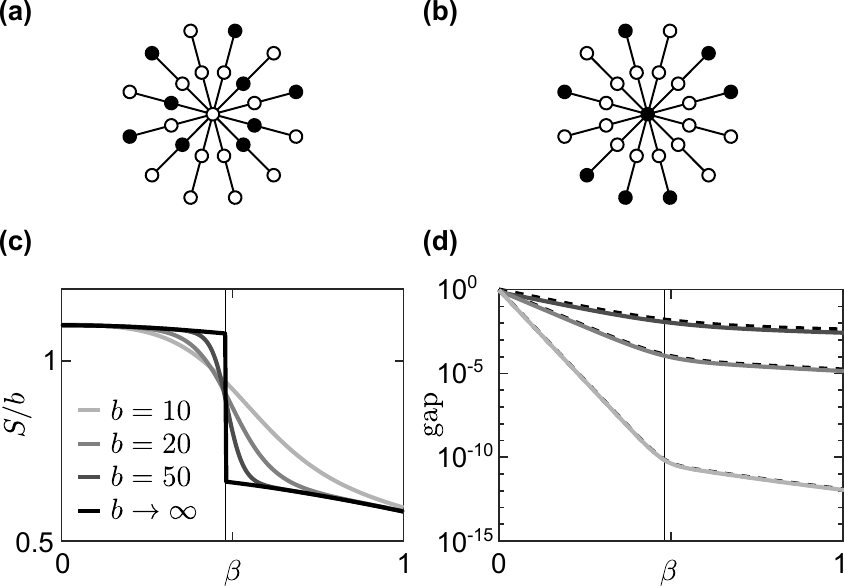}
  \caption{Sampling from weighted independent sets of a star graph. (a), (b)~Examples of independent sets of a star graph with $b$ branches and two vertices per branch. The weight of the central vertex is $b$, while all other vertices are assigned weight $1$. The filled vertices indicate members of an independent set (occupied vertices). (c)~Classical entropy of the Gibbs distribution of weighted independent sets of the star graph as a function of temperature for various system sizes. There is a discontinuous phase transition at $\beta^* \approx 0.48$. (d)~Gap of the parent Hamiltonian for the same system sizes as in (c), excluding $b \to \infty$. The solid curves are numerically exact results for the gap in the completely symmetric subspace, while the dashed curves are obtained using the approximate two-state model.}
  \label{fig:fig8}
\end{figure}

In the previous examples, sampling becomes hard only at zero temperature. However, sampling from the zero-temperature Gibbs distribution is equivalent to minimizing the energy of the classical Hamiltonian, for which there may exists more efficient algorithms than Markov chain Monte Carlo. It is therefore of interest to analyze a sampling problem with a Markov chain that mixes slowly at finite temperature. To this end, we consider sampling from independent sets of the star graph shown in \figref{fig:fig8}(a) and (b). The star graph consists of a central vertex and $b$ branches with two vertices per branch. The independent sets are weighted with a weight $w_c = b$ for the central vertex and $w_i = 1$ for all other vertices. Even though the star graph is not a unit disk graph, it may be possible to implement the corresponding parent Hamiltonian using anisotropic Rydberg interactions~\cite{Glaetzle2014}.

Before discussing the Markov chain and the quantum dynamics of this model, it is helpful to consider the classical equilibrium phase diagram. The partition function corresponding to the classical Hamiltonian $H_c$ in \eqref{eq:hc_wis} is given by $\mathcal{Z} = \left(1 + 2 e^{\beta}\right)^b + e^{b \beta} \left(1 + e^{\beta}\right)^b$. The two terms arise from the different configurations of the central vertex. From the Helmholtz free energy $F = - \log \mathcal{Z} / \beta$ and the total energy $U = - \partial \log \mathcal{Z} / \partial \beta$, the entropy can be computed as $S = \beta (U - F)$, which is plotted in \figref{fig:fig8}(c). In the thermodynamic limit, the system undergoes a discontinuous phase transition at $\beta^* = \log \varphi$, where $\varphi = (\sqrt{5} + 1)/2$ is the golden ratio. The origin of the phase transition can be understood by noting that the probability that the central site is occupied is given by
\begin{equation}
  p_1 = \left[ 1 + e^{-b \beta} \left( \frac{1 + 2 e^\beta}{1  + e^{\beta}} \right)^b \right]^{-1}.
\end{equation}
This expression turns into the step function $p_1 = \Theta(\beta - \beta^*)$ in the thermodynamic limit $b \to \infty$. At high temperature, it is entropically favorable for the central vertex to be unoccupied as this allows the branches to be in $3^b$ distinct configurations. Below the phase transition temperature, the reduction in energy from occupying the central vertex outweighs the entropic cost of reducing the number of configurations to $2^b$ due to the independent set constraint.

The Markov chain with single spin flips on this graph is subject to a severe kinetic obstruction since changing the central vertex from unoccupied to occupied requires all neighboring vertices to be unoccupied. Assuming that each individual branch is in thermal equilibrium, the probability of accepting such a move is given by $p_{0 \to 1} = [(1 + e^{\beta})/(1 + 2 e^{\beta})]^{b}$. The reverse process is energetically suppressed with an acceptance probability $p_{1 \to 0} = e^{-b \beta}$. The central vertex can thus become trapped in the thermodynamically unfavorable configuration, resulting in a mixing time that grows exponentially with the number of branches $b$ at any finite temperature. Indeed, we will see below that the spectral gap of the parent Hamiltonian (and thus the Markov chain) is approximately equal to $p_{1 \to 0}$ for temperatures above the phase transition and to $p_{0 \to 1}$ below. We remark that despite the exponentially large mixing time, it is possible to sample efficiently at high temperature if the initial configuration for the Markov chain is drawn from the uniform distribution. Since the probability of the central vertex being initially occupied is exponentially small, the desired Gibbs distribution can be approximated with exponentially small error by restarting the Markov chain multiple times.  By the same argument, the Markov chain almost certainly starts in the wrong configuration in the low temperature phase and the exponentially large mixing time $t_m \gtrsim 1/p_{0 \to 1}$ is a limiting factor.

To determine the quantum phase diagram associated with the parent Hamiltonian \eqref{eq:parent_wis} for the star graph, we observe that the parent Hamiltonian is invariant under permutations of the branches. Restricting the analysis to the completely symmetric subspace, we show in \appref{app:star} that $H_q(\beta)$ possesses a two-dimensional low-energy subspace. The subspace is to an excellent approximation spanned by the states $\ket{\psi_0(\beta)}$ and $\ket{\psi_1(\beta)}$, which represent the Gibbs states with the central vertex respectively fixed to be unoccupied or occupied. This is true even when the parameters $\Omega_i$ and $V_{e,i}$ associated with the central vertex, denoted by $\Omega_\mathrm{cen}$ and $V_{e, \mathrm{cen}}$, are varied arbitrarily, assuming all other terms in the Hamiltonian follow the one-parameter family $H_q(\beta)$. The effective Hamiltonian in the low-energy subspace can be obtained by a Schrieffer--Wolff transformation~\cite{Bravyi2011}, which to second order yields
\begin{equation}
  H_\mathrm{eff} = 
  \begin{pmatrix}
    (1 - f) \varepsilon_0 & -( 1 - f) J\\
    -(1- f) J & V_{e,\mathrm{cen}} - f \Omega_\mathrm{cen}^2
  \end{pmatrix},
  \label{eq:h_eff}
\end{equation}
where
\begin{gather}
  \varepsilon_0 = \left( \frac{1 + e^{\beta}}{1 + 2 e^{\beta}} \right)^b,  \qquad
  J = \Omega_\mathrm{cen}\left( \frac{1 + e^{\beta}}{1 + 2 e^{\beta}} \right)^{b/2}.
  \label{eq:h_eff_params}
\end{gather}
The term $f$ is a second order correction, which vanishes as $1/b$ in the thermodynamic limit.

Since $\ket{\psi_0(\beta)}$ and $\ket{\psi_1(\beta)}$ represent macroscopically distinct states, the system undergoes a first-order quantum phase transition when the bias $(1 - f) \varepsilon_0 - (V_{e, \mathrm{cen}} - f \Omega_\mathrm{cen}^2)$ vanishes. Along the one-parameter family $H_q(\beta)$, we have $V_{e, \mathrm{cen}} = \Omega_\mathrm{cen}^2$ such that the phase transition occurs when $\varepsilon_0 = V_{e,\mathrm{cen}}$. Solving for $\beta$ gives the critical value $\beta^* = \log \varphi$, in agreement with the location of the classical phase transition. More generally, \eqref{eq:h_eff} predicts a gap given by $(1 - f) [ e^{- b \beta} + (1 + e^\beta)^b/(1 + 2 e^\beta)^b]$. As shown in \figref{fig:fig8}(d), this expression agrees well with the numerically exact result of the gap in the permutation symmetric subspace for large values $b$. The two-state model thus tells us that the gap vanishes as $e^{-b \beta}$ above the phase transition ($\beta < \beta^*$) and as $(1 + e^{\beta})^b/(1 + 2 e^{\beta})^b$ below ($\beta > \beta^*$) up to a small correction due to the $(1 - f)$ prefactor. This result perfectly matches the estimates of the mixing time of the Markov chain based on the transition probabilities $p_{0 \to 1}$ and $p_{1 \to 0}$. It is worth pointing out that the location of the quantum phase transition is well defined even though the gap vanishes for any $\beta > 0$ in the thermodynamic limit. We can define the critical region of the phase transition as the range of parameters for which the tunneling rate $(1 - f) J$ exceeds the bias $(1 - f) \varepsilon_0 - (V_{e, \mathrm{cen}} - f \Omega_\mathrm{cen}^2)$. The critical region is of size $\Delta \beta \sim 1/b$, such that the phase transition point $\beta^*$ is indeed meaningful.

Having described the Markov chain and the related spectral properties of the parent Hamiltonian $H_q(\beta)$, we proceed to the adiabatic state preparation of the Gibbs state. We note that the Gibbs state at $\beta = 0$ can be efficiently prepared from a product state. For instance, one can start with $V_{e,i} = 1$ and $V_{g,i} = \Omega_i = 0$, where the ground state is the state of all vertices unoccupied. Next, all $\Omega_i$ are simultaneously ramped up to $1$ at a constant rate before doing the same for $V_{g,i}$. The Hamiltonian remains gapped along this path such that the adiabatic state preparation proceeds efficiently. Hence, we may use the Gibbs state at $\beta = 0$ as the initial point for all adiabatic paths. Despite the exponentially small gap, it is also possible to efficiently prepare any state above the phase transition. The probability of occupying the central vertex is exponentially small above the critical temperature, outside the critical region. Therefore, an exponentially good approximation to the Gibbs state can be prepared along the one-parameter family $H_q(\beta)$ since adjusting the amplitudes of the different configurations of the branches proceeds without kinetic obstruction.

\begin{figure}[t]
  \centering
  \includegraphics{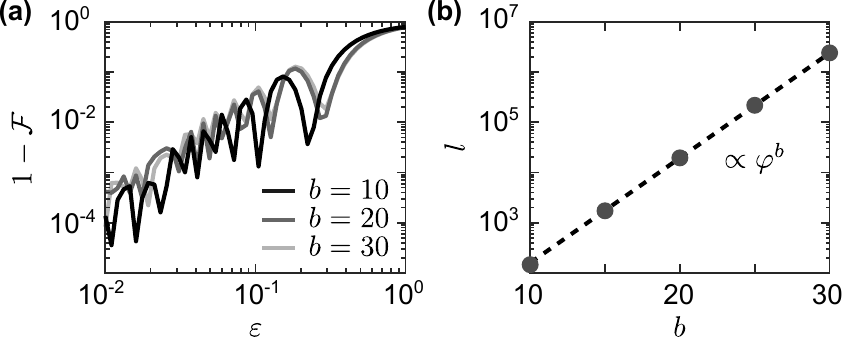}
  \caption{Adiabatic state preparation of $\ket{\psi(2 \beta^*)}$ for the star graph along the one-parameter family $H_q(\beta)$ staring from $\beta = 0$. (a)~Infidelity $1 - \mathcal{F}$ as a function of $\varepsilon$ for different numbers of branches $b$. (b)~Adiabatic path length $l$ as a function of $b$. Since the infidelity in (a) is largely independent of $b$, the adiabatic state preparation time $t_a$ is approximately proportional to $l$. The numerically exact result for the adiabatic path length (dots) agrees well with the $l \propto \varphi^b$ dependence (dashed line) expected from the tunneling rate.}
  \label{fig:fig9}
\end{figure}

As for the Markov chain, this line of reasoning fails when the target temperature is below the phase transition.  Along the one-parameter family $H_q(\beta)$, the gap at the phase transition point is equal to the tunneling rate $J = \varphi^{-b}$, ignoring the subleading factor $(1-f)$. Hence, we expect the adiabatic state preparation time to be proportional to $\varphi^b$, which we numerically confirmed for the final state $\ket{\psi(2 \beta^*)}$ in \figref{fig:fig9}. The numerical integration of the Schrödinger equation follows the same steps as for the independent set problem on the chain graph (see \appref{app:wis_numerical}) with time steps chosen to satisfy $\Delta t || \di \ket{0} / \di t|| = 10^{-4}$. We note that the lower bound on the mixing time of the Markov chain at the phase transition has the same scaling $t_m \gtrsim \varphi^b$. For sampling below the phase transition, the mixing time increases such that the quantum algorithm appears to achieve a speedup. However, the slowdown of the Markov chain can be circumvented by performing simulated annealing~\cite{Kirkpatrick1983} across the phase transition. While a detailed analysis of simulated annealing is beyond the scope of the present work, we expect that its convergence time matches the time to adiabatically prepare a state along the one-parameter family $H_q(\beta)$ for any final value of $\beta$ since the occupation of the central vertex is essentially frozen outside the critical region of the phase transition.

The adiabatic state preparation time along the one-parameter family $H_q(\beta)$ is limited by the tunneling rate $J$ at the phase transition. Since $J$ is proportional to $\Omega_\mathrm{cen}$, it is natural to consider trajectories along which $\Omega_\mathrm{cen}$ is of order one at the phase transition rather than $e^{-b \beta^*/2}$. Because $[(1 + e^{\beta^*})/(1 + 2 e^{\beta^*})]^b = e^{-b \beta^*}$, this simple argument together with \eqref{eq:h_eff_params} predicts a quadratic speedup.  As a first guess, one may consider a path where $\Omega_\mathrm{cen}$ is held constant at $1$, while all other parameters are varied according to $H_q(\beta)$ from $\beta = 0$ to the desired final value of $\beta$, taken to be $2 \beta^*$ in what follows. The time required to cross the phase transition is again $t_a \sim 1/J$, where $J$ is evaluated at the phase transition. The term $f \Omega_\mathrm{cen}^2$ in the effective Hamiltonian \eqref{eq:h_eff} shifts the phase transition close to $\beta = 0$ such that $t_a \sim (3/2)^{b/2}$. However, this does not yet result in a speedup for preparing the Gibbs state as one still has to decrease $\Omega_\mathrm{cen}$ to its final value $e^{- b \beta^*}$. It turns out that this step negates the speedup if performed adiabatically. The reason for this is that the large value of $\Omega_\mathrm{cen}$ admixes states where the central vertex is unoccupied, skewing the occupation probability away from its thermal expectation value. Even though this admixture is small, on the order of $f$, the time to adiabatically remove it exceeds the time to initially cross the phase transition. To avoid this issue, one can in principle suddenly switch $\Omega_\mathrm{cen}$ to its final value at the cost that the final fidelity is limited by the admixture. Perturbation theory and numerical results indicate that this results in an infidelity $1 - \mathcal{F}$ that decays as $1/b^2$ for large values of $b$.
\begin{figure}[t]
  \centering
  \includegraphics{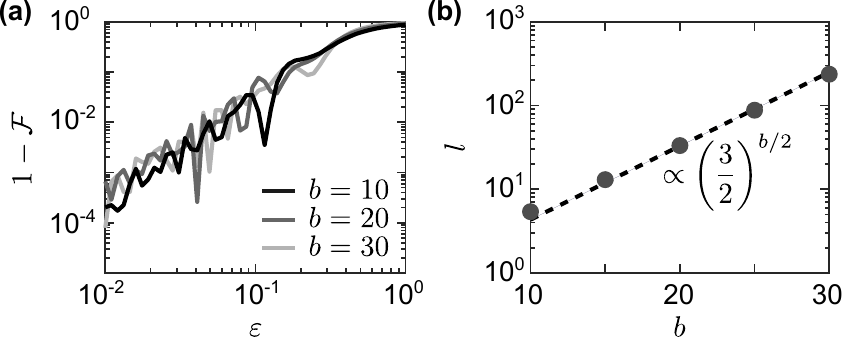}
  \caption{Adiabatic state preparation of $\ket{\psi(2 \beta^*)}$ for the star graph along an improved path. The path starts at $H_q(0)$ before decreasing $V_{e,\mathrm{cen}}$ from $1$ to $-1$. All other parameters are subsequently varied according to the one-parameter family $H_q(\beta)$ from $\beta = 0$ to $\beta = 2 \beta^*$. Note that the final point differs from $H_q(2\beta^*)$ by the value of $V_{e,\mathrm{cen}}$. (a)~Infidelity $1 - \mathcal{F}$ as a function of $\varepsilon$ for different numbers of branches $b$. (b)~Adiabatic path length $l$ as a function of $b$. Since the infidelity in (a) is largely independent of $b$, the adiabatic state preparation time $t_a$ is approximately proportional to $l$. The dots show the numerically exact results, while the dashed line is a guide to the eye displaying the exponential function $l \propto (3/2)^{b/2}$.}
  \label{fig:fig10}
\end{figure}

\begin{figure}[t]
  \centering
  \includegraphics{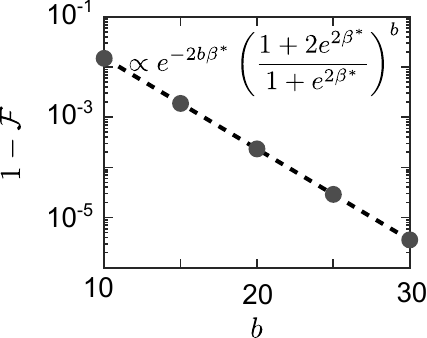}
  \caption{Final step in the state preparation for the star graph, following the adiabatic sweep in \figref{fig:fig10}. The value of $V_{e,\mathrm{cen}}$ is switched suddenly from $-1$ to $e^{- 2 b \beta^*}$. The dots show the resulting infidelity, assuming perfect fidelity along the preceding adiabatic path. The infidelity agrees well with the probability that the central vertex is unoccupied (dashed line, whose functional form is shown in the plot).}
  \label{fig:fig11}
\end{figure}

A slight modification of the path achieves a final state infidelity that is not only polynomially but exponentially small in $b$. First, $V_{e, \mathrm{cen}}$ is lowered from its initial value~$1$ to $-1$, which can be done in time $t_a \sim (3/2)^{b/2}$ as before. Next, all other parameters are varied along $H_q(\beta)$, for which only a time polynomial in $b$ is required. Numerical results for these two steps are shown in \figref{fig:fig10}. Finally, $V_{e, \mathrm{cen}}$ is ramped to its final value $e^{-2 b \beta^*}$. Similar to the previous scheme, perfectly adiabatic evolution of this last step would require a very long time because the probability that the central vertex is unoccupied is initially too small due to the large, negative value of $V_{e, \mathrm{cen}}$. However, since its final value $p_0 \approx e^{-2 b \beta^*} [(1 + 2 e^{2 \beta^*})/(1+e^{2 \beta^*})]^b$ is exponentially small in $b$, the fidelity can be exponentially close to unity when suddenly switching $V_{e,\mathrm{cen}}$ to its final value. We numerically confirmed this prediction for the sudden transition in \figref{fig:fig11}.

To summarize, we identified a suitable path along which the adiabatic evolution, supplemented by a sudden change of parameters at the end, achieves a preparation time of $t_a \sim (3/2)^{b/2}$. The speedup over the Markov chain with mixing time $t_m \gtrsim \varphi^b$ at the phase transition appears to be more than quadratic. However, it is likely that a classical computation time on the order of $(3/2)^b$ can be achieved using simulated annealing. For instance, one could consider an annealing schedule in which the weight on the central vertex is first increased. This shifts the phase transition towards $\beta = 0$, allowing one to sample at the phase transition in a time that scales as $(3/2)^b$. In the annealing schedule, the temperature can then be lowered to the desired value before ramping the weight of the central vertex back to its initial value. This annealing schedule is in many ways similar to the adiabatic path discussed above. It is nevertheless quadratically slower because unlike in the quantum case it is not possible to vary $V_{e, \mathrm{cen}}$ and $\Omega_\mathrm{cen}$ independently.

The above results for the star graph exhibit important differences from the chain graph and from sampling from the Ising model. While the speedup is also quadratic, it originates from coherent tunneling as opposed to the ballistic propagation of domain walls. This is related to the fact that the parent Hamiltonian for the star graph exhibits a first-order quantum phase transition as opposed to the second-order transition in the preceding models. In the next section, we discuss another model in which the dynamics are governed by a first-order transition.

\section{Unstructured search problem\label{sec:search}}

The unstructured search problem was pivotal in the development of quantum algorithms. Grover's algorithm gave an early example of a provable quantum speedup and it remains an essential subroutine in many proposed quantum algorithms~\cite{Grover1996}. Moreover, the unstructured search problem played a crucial role in the conception of adiabatic quantum computing~\cite{Roland2002}. Below, we show that when applied to the unstructured search problem, our formalism recovers the adiabatic quantum search algorithm along with its quadratic speedup over any classical algorithm. While the nonlocality of the resulting parent Hamiltonian render it challenging to implement in practice, the result underlines the power of our approach in enabling quantum speedup in a setting where the speedup is provably optimal~\cite{Zalka1999}.

We consider the problem of identifying a single marked configuration $m$ in a space of a total of $N$ elements.  To connect this search problem to a sampling problem, we assign the energy $-1$ to the marked configuration, while all other states have energy $0$. This is summarized by the classical Hamiltonian
\begin{equation}
  H_c = - \ket{m} \bra{m}.
\end{equation}
Solving the search problem may now be formulated as sampling from the Gibbs distribution associated with $H_c$ at zero temperature. Given the lack of structure of the problem, a natural choice for the Markov chain is to propose any configuration with equal probability $1/N$~\footnote{Other updates such as single-spin flips when the configurations are encoded into $n = \log N$ bits can also lead to a quadratic speedup.}. If the update is accepted according to the Metropolis--Hastings rule, the resulting parent Hamiltonian takes the form
\begin{align}
  H_q(\beta) &= \mathbb{I} - A(\beta) \left( \ket{m} \bra{m} + \ket{m_\perp} \bra{m_\perp} \right) \nonumber \\
  & \hspace{1cm} - V_0(\beta) \ket{\psi_0} \bra{\psi_0} - V_m(\beta) \ket{m} \bra{m},
  \label{eq:parent_search}
\end{align}
where $A(\beta) = (N-1) \left( 1 - e^{- \beta / 2} \right)/N$, $V_0(\beta) = e^{- \beta / 2}$, and $V_m(\beta) = [1 + (N-2) e^{- \beta / 2} - (N-1) e^{-\beta}]/N$.
The states $\ket{\psi_0} = \sum_{i} \ket{i} / \sqrt{N}$ and $\ket{m_\perp} = \sum_{i \neq m} \ket{i} / \sqrt{N-1} $ are equal superpositions of all states in the search space with and without the marked state $\ket{m}$. For conciseness, we have omitted the factor $n = \log N$ that would render the parent Hamiltonian extensive from \eqref{eq:parent_search} as it represents only a logarithmic correction to the computation time.

Since $\ket{\psi_0}$ is contained in the subspace spanned by $\{\ket{m}, \ket{m_\perp}\}$, the Hamiltonian acts trivially on the orthogonal subspace. In fact, all nontrivial dynamics arise from the second line of \eqref{eq:parent_search}. Considering the extended parameter space of arbitrary $(V_0, V_m)$, we can show that the Hamiltonian supports two distinct quantum phases separated by a first-order phase transition. By rewriting \eqref{eq:parent_search} in terms of $\ket{m}$ and $\ket{m_\perp}$ only, one can show that these two states have a relative energy difference $V_m - (1 - 2/N)V_0$ while being connected by an off-diagonal matrix element (tunneling rate) of magnitude $\sqrt{N-1} V_0/N$. In the thermodynamic limit $N \to \infty$, the system thus undergoes a first-order quantum phase transition when $V_m = V_0$. For $V_m > V_0$, the ground state has large overlap with $\ket{m}$, whereas for $V_m < V_0$, it is close to $\ket{m_\perp}$ (and $\ket{\psi_0}$).

\begin{figure}[t]
  \centering
  \includegraphics{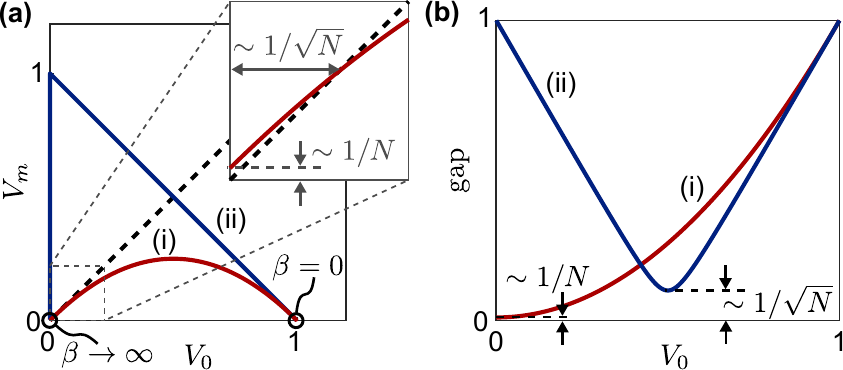}
  \caption{Unstructured search algorithm. (a) Two-dimensional parameter space corresponding to the Hamiltonian in \eqref{eq:parent_search}, where the dashed line indicates the location of a first-order quantum phase transition. The red curve, path (i), shows the one-parameter family $H_q(\beta)$. The blue path (ii) is equivalent to the adiabatic quantum search algorithm. The inset displays a magnified view of a region of parameter space close to the origin. (b) The gap of the Hamiltonian in \eqref{eq:parent_search} along paths (i) and (ii) in panel (a).}
  \label{fig:fig12}
\end{figure}

The one-parameter family $H_q(\beta)$, indicated by the red curve (i) in \figref{fig:fig12}(a), traces out a segment of a parabola passing through $(V_0, V_m) = (1, 0)$ when $\beta = 0$ and $(V_0, V_m) = (0, 1/N)$ as $\beta \to \infty$. The gap at zero temperature is equal to $1/N$ [\figref{fig:fig12}(b)], which allows us to bound the mixing time by $t_m \gtrsim N$ (up to logarithmic corrections). This bound is expected as any classical algorithm must check on average half the configurations to solve the unstructured search problem. As in all previous examples, adiabatic state preparation along the one-parameter family leads to the same time complexity. To see this, we assume that the ground state at $\beta = 0$, given by $\ket{\psi_0}$, can be readily prepared. Adiabatic state preparation experiences a bottleneck close to the quantum phase transition, where $V_0$ and $V_m$ are on the order of $1/\sqrt{N}$ [see inset of \figref{fig:fig12}(a)]. Since the adiabatic state preparation time is limited by the inverse of the tunneling rate in the critical region, we obtain an adiabatic state preparation time $t_a \sim \sqrt{N}/V_0 \sim N$.

Similar to what we found for the star graph, the adiabatic state preparation can be sped up by crossing the phase transition at a point where the tunneling rate is large. In particular, this requires that $V_0$ and $V_m$ be of order one. One such path is path (ii) shown in \figref{fig:fig12}(a). A straight line segment connects $(V_0, V_m) = (1,0)$ to $(0, 1)$ before continuing to the final point $(V_0, V_m) = (0, 1/N)$. Since the Hamiltonian is purely diagonal along $V_0 = 0$ in the computational basis, there are no diabatic transitions along the latter segment and the parameters can be changed suddenly. The gap along the former segment is shown in \figref{fig:fig12}(b). The corresponding Hamiltonian is in fact identical to the Hamiltonian of the adiabatic quantum search algorithm, which was derived by interpolating between the projectors $\ket{\psi_0} \bra{\psi_0}$ and $\ket{m} \bra{m}$~\cite{Roland2002}. It was shown that by carefully choosing the rate of change using a scheme essentially equivalent to that outlined in \secref{sec:adiabatic}, it is indeed possible to prepare the final ground state with high fidelity in a time $t_a \sim \sqrt{N}$ limited by the tunneling rate at the phase transition.

\section{Summary and outlook\label{sec:summary}}
In summary, we have described a method to construct quantum algorithms to sample from Gibbs distributions. The approach can be readily generalized to any probability distribution that can be described as the stationary distribution of a Markov chain satisfying detailed balance. Our results differ from previous work by considering adiabatic state preparation in a parameter space that has been extended beyond the one-parameter family of Hamiltonians $H_q(\beta)$. By means of four examples, we showed that  it is possible to achieve a quantum speedup by suitably navigating the quantum phases in the extended parameter space. The speedup has a different origin depending on the nature of the phase transition. In the case of second-order phase transitions, the speedup was due to the ballistic propagation of domain walls as opposed to diffusive motion in the classical Markov chain. For first-order phase transitions, we could trace the speedup to coherent tunneling between macroscopically distinct states.

The quantum Hamiltonians encountered in our construction are guaranteed to be local provided that the Gibbs distribution originates from a local classical Hamiltonian and that the Markov chain updates are local. This was the case in all of the examples except for the unstructured search problem, which we included to highlight the power of the approach in a more abstract setting. It is therefore possible to efficiently implement time evolution under these Hamiltonians on a universal quantum computer using Hamiltonian simulation~\cite{Lloyd1996}. Moreover, for sampling from independent sets of unit disk graphs, there exists a natural implementation of the parent Hamiltonian using Rydberg states of neutral atoms. The proposed scheme is compatible with existing architectures~\cite{Labuhn2016,Bernien2017,Ebadi2020,Scholl2020}, opening the door to exploration of sampling problems on near-term quantum devices.

Further work is required to extend the applicability of our approach to a wider range problems. As a first step, one may consider a generalization of the above spin models to higher dimensions. For instance, Glauber dynamics in the two-dimensional Ising model differs substantially from its one-dimensional counterpart owing to presence of a finite-temperature phase transition in the classical model. At temperatures above the phase transition, the Markov chain still mixes rapidly, whereas at low temperature, the mixing time diverges exponentially with the linear dimension of the system~{\cite{Lubetzky2012}}. Moreover, there exist many configurations with large domains, which relax slowly to equilibrium in the Markov chain. Below the phase transition, the corresponding parent Hamiltonian $H_q(\beta)$ therefore describes a a system with a large number of states energetically close to the ground state, hinting at the presence of an unusual quantum phase. Additional research is needed to fully characterize this quantum phase, e.g.~using quantum Monte Carlo techniques, and to identify possible mechanisms for quantum speedup when adiabatically approaching it.

Practically relevant, hard sampling problems, such as sampling from the Gibbs distribution of a classical spin glass or other disordered models in two or more dimensions, lack much of the structure such as translational symmetry of the problems discussed above. Here, a key challenge is to identify suitable adiabatic paths as it may not be possible to determine the complete phase diagram. This issue could potentially be addressed by employing hybrid algorithms which combine quantum evolution with classical optimization to identify a good adiabatic path~{\cite{Schiffer2021}}. More concretely, the energy of the parent Hamiltonian $H_q(\beta)$ could be minimized using variational quantum algorithms similar to existing proposals but without the need for complex measurements of the entanglement entropy~\cite{Wu2019,Martyn2019}. Since variational algorithms do not require a physical realization of the parent Hamiltonian, this approach could be particularly fruitful for complex cost functions composed of several parent Hamiltonians involving nonlocal cluster updates such as those of the Swendsen--Wang algorithm~\cite{Swendsen1987}. 

\begin{acknowledgments}
    We thank J. I. Cirac, E. A. Demler, E. Farhi, A. Polkovnikov, and P. Zoller for insightful discussions. This work was supported by the National Science Foudation, the MIT–Harvard Center for Ultracold Atoms,the Department of Energy, and the DARPA ONISQ program. D.S. was supported by AFOSR: Grant FA9550-21-1-0236. The Flatiron Institute is a division of the Simons Foundation.
\end{acknowledgments}

\appendix
\section{Ising chain}
\subsection{Free-fermion solution\label{app:fermion}}
The Hamiltonian in \eqref{eq:parent_ising} can be mapped onto a free-fermion model using a Jordan--Wigner transformation. We define the fermion annihilation and creation operators $a_i$, $a_i^\dagger$ and relate them to the Pauli matrices according to
\begin{align}
  \sigma_i^x &= 2 a_i^\dagger a_i - 1, \\
  \frac{1}{2} \left( \sigma_i^z + i \sigma_i^y \right) &= e^{i \pi \sum_{j=1}^{i-1} a_j^\dagger a_j} a_i, \\
  \frac{1}{2} \left( \sigma_i^z - i \sigma_i^y \right) &= e^{i \pi \sum_{j=1}^{i-1} a_j^\dagger a_j} a_i^\dagger.
\end{align}
\Eqref{eq:parent_ising} becomes, up to a constant,
\begin{align}
  \label{eq:boundary}
  H_q = &- h \sum_{i = 1}^n (2 a_i^\dagger a_i - 1) - J_1 \sum_{i = 1}^{n-1} ( a_i^\dagger - a_i ) ( a_{i+1}^\dagger + a_{i+1} ) \nonumber\\
  &- J_2 \sum_{i = 1}^{n-2} ( a_{i}^\dagger - a_{i} ) ( a_{i+2}^\dagger + a_{i+2} ) \nonumber\\
  &+ e^{i \pi N}  \left[ J_1 (a_n^\dagger - a_n)(a_1^\dagger + a_1) \right. \nonumber\\
    &\hspace{2cm}+ J_2 (a_{n-1}^\dagger - a_{n-1}) (a_1^\dagger + a_1) \nonumber\\
  &\hspace{2cm}+ \left. J_2 (a_{n}^\dagger - a_{n}) (a_2^\dagger + a_2) \right], 
\end{align}
where $N = \sum_{i=1}^n a_i^\dagger a_i$ is the total number of fermions. While the fermion number itself is not conserved, the parity $e^{i \pi N}$ is, allowing us to consider the even and odd subspaces independently. 

We define the momentum space operators
\begin{equation}
  a_k = \frac{1}{\sqrt{n}} \sum_{j = 1}^n e^{-i k j} a_j,
\end{equation}
which satisfy fermionic commutation relations for suitably chosen $k$. We let
\begin{equation}
  k = \frac{2 \pi}{n} \times
  \begin{cases}
    (l + 1/2) & \text{if $N$ is even}\\
    l & \text{if $N$ is odd}
  \end{cases}
\end{equation}
for $l = 0, 1, \ldots, n-1 \,\, (\mathrm{mod} \, n)$. With this definition, the inverse Fourier transformed operators have the formal property $a_{i + n} = - e^{i \pi N} a_i$, which accounts for the boundary terms in \eqref{eq:boundary}. The Hamiltonian simplifies to
\begin{equation}
  H_q = \sum_k 
  \begin{pmatrix}
    a_k^\dagger, & a_{-k}
  \end{pmatrix}
  \begin{pmatrix}
    A_k & - i B_k\\
    i B_k & - A_k
  \end{pmatrix}
  \begin{pmatrix}
    a_k\\
    a_{-k}^\dagger
  \end{pmatrix}.
\end{equation}
$A_k = -h - J_1 \cos (k) - J_2 \cos (2k)$, $B_k = J_1 \sin (k) + J_2 \sin (2k)$
While the above Hamiltonian can be diagonalized by a standard Bogoliubov transformation, it will prove more convenient for our purposes to map it onto noninteracting spins. For $0 < k < \pi$, we define
\begin{align}
  \tau_k^x &= a_k^\dagger a_{-k}^\dagger + a_{-k} a_k, \\
  \tau_k^y &= - i (a_k^\dagger a_{-k}^\dagger - a_{-k} a_k ), \\
  \tau_k^z &= a_k^\dagger a_{k} - a_{-k} a_{-k}^\dagger.
\end{align}
It is straightforward to check that these operators satisfy the same commutation relations as Pauli matrices. In addition, operators corresponding to different values of $k$ commute such that we can view them as independent spin-$1/2$ systems, one for each value of $k$. We restrict the range of momenta to $0 < k < \pi$ due to the redundancy $\tau_{-k}^\alpha = - \tau_k^\alpha$. The cases $k = 0$ and $k = \pi$ require special treatment as both $\tau_k^x$ and $\tau_k^y$ vanish.

For concreteness, we assume that the number of spins $n$ is even. The special cases $k = 0$ and $k = \pi$ are then both part of the odd parity subspace ($e^{i \pi N} = -1$). The Hamiltonian of the even parity subspace can be written as
\begin{equation}
  H_q^\mathrm{even} = 2 \sum_{0 < k < \pi} E_k
  \left( \cos \theta_k \tau_k^z + \sin \theta_k \tau_k^y \right),
  \label{eq:spins}
\end{equation}
where
\begin{equation}
  E_k = \sqrt{ A_k^2 + B_k^2 }.
  \label{eq:energy}
\end{equation}
The angles $\theta_k$ are uniquely defined by
\begin{align}
  E_k \cos \theta_k &= A_k, \\
  E_k \sin \theta_k &= B_k.
  \label{eq:theta}
\end{align}
The ground state is given by
\begin{equation}
  \ket{0}_\mathrm{even} = \prod_{0 < k < \pi} e^{i \theta_k \tau_k^x/2} \ket{\mathrm{vac}},
  \label{eq:gs}
\end{equation}
where $\ket{\mathrm{vac}}$ is the vacuum with respect to the $a_k$ operators. The ground state energy is
\begin{equation}
  E_0^\mathrm{even} = - 2 \sum_{0 < k < \pi} E_k.
\end{equation}

In the odd parity subspace, we have
\begin{align}
  H_q^\mathrm{odd} = 2 \sum_{0 < k < \pi} &E_k \left( \cos \theta_k \tau_k^z + \sin \theta_k \tau_k^y \right) \nonumber\\
  &- (h + J_1 + J_2) (2 a_0^\dagger a_0 - 1) \nonumber\\
  &- (h - J_1 + J_2) (2 a_\pi^\dagger a_\pi - 1).
\end{align}
The construction of the ground state is analogous to the even case with the additional requirement that either the $a_0$ fermion or the $a_\pi$ fermion, whichever has the lower energy, be occupied. One can show that the resulting energy is gapped above $E_0^\mathrm{even}$ when $h + J_1 + J_2$ and $h - J_1 + J_2$ have the same sign. In the case of opposite signs, the even and odd sector ground states are degenerate in the thermodynamic limit, corresponding to the symmetry breaking ground states of the ferromagnetic phase.

In the main text, we consider adiabatic evolution starting from the ground state at $J_1 = J_2 = 0$. Following the above discussion, this state is part of the even subspace. Since the time evolution preserves parity, we may restrict our discussion to the even subspace, dropping all associated labels in what follows.

Excited states can be constructed by flipping any of the $\tau$ spins. Since any spin rotation commutes with the parity operator, singly excited states are given by
\begin{equation}
  \ket{k} = \tau_k^x \ket{0},
  \label{eq:es}
\end{equation}
with an energy $4 E_k$ above the ground state. 

\subsection{Time-dependent Schrödinger equation\label{app:tdse}}
To compute the fidelity, we numerically integrate the Schrödinger equation for each spin $\tau_k$. We work in the instantaneous eigenbasis $\ket{\chi_k^\pm(t)}$, which are eigenstates of $H_k = 2 E_k \left(\cos \theta_k \tau_k^z + \sin \theta_k \tau_k^y \right)$ with energies $\pm 2 E_k$ [see \eqref{eq:spins}]. It is convenient to parametrize each adiabatic path by a dimensionless time $s$ running from $0$ to $1$. Writing the state at time $s$ as
\begin{equation}
  \ket{\psi_k(s)} = c_k(s) \ket{\chi_{k}^-(s)} + d_k(s) \ket{\chi_k^+(s)},
\end{equation}
the coefficient $c_k$ and $d_k$ are determined by the Schrödinger equation
\begin{equation}
  i \frac{\di}{\di s}
  \begin{pmatrix}
    c_k\\
    d_k
  \end{pmatrix}
  = 
  \begin{pmatrix}
    -2 E_k(s) \frac{\di t}{\di s} & - \frac{i}{2} \frac{\di \theta_k}{\di s}\\
    \frac{i}{2} \frac{\di \theta_k}{\di s} & 2 E_k(s) \frac{\di t}{\di s}
  \end{pmatrix}
  \begin{pmatrix}
    c_k\\
    d_k
  \end{pmatrix}
  \label{eq:schroedinger}
\end{equation}
with the initial condition $c_k(0) = 1$, $d_k(0) = 0$.  The final fidelity is obtained by solving this equation for each spin and multiplying the individual fidelities,
\begin{equation}
  \mathcal{F} = \prod_{0 < k < \pi} |c_k(1)|^2.
\end{equation}

We note that all terms in \eqref{eq:schroedinger} can be evaluated without having to solve for the physical evolution time $t(s)$. The terms $E_k(s)$ and $\di \theta_k / \di s$ are readily computed from equations~(\ref{eq:energy})--(\ref{eq:theta}), while $\di t / \di s$ follows from equation (5) of the main text:
\begin{equation}
  \frac{\di t}{\di s} = \frac{1}{\varepsilon} \sqrt{\sum_{\mu, \nu} g_{\mu \nu}(s) \frac{\di \lambda_\mu}{\di s} \frac{\di \lambda_\nu}{\di s}}.
\end{equation}
Here, $\lambda_1 = J_1$, $\lambda_2 = J_2$, setting $h = 1$ throughout. To vary the total evolution time $t_\mathrm{tot}$, we simply adjust the value of $\varepsilon$.
We obtained good convergence by evolving under constant $s = s_n$ for an interval $\Delta s_n = 2 \times 10^{-3} / \left|\frac{\di \theta_k}{\di s} \right|_{s = s_n}$ before incrementing $s_{n+1} = s_n + \Delta s_n$. The number of steps is independent of the total time, yet the final fidelity is well estimated since the probability of leaving the ground state is small in each step.  

\subsection{Adiabatic path length\label{app:l}}
To compute the adiabatic path length, we note that it follows from \eqref{eq:gs} and \eqref{eq:es} that
\begin{equation}
  \partial_\mu \ket{0} = \frac{i}{2} \sum_{0 < k < \pi} \partial_\mu \theta_k \ket{k}.
\end{equation}
From the definition of $g_{\mu \nu}$ in \eqref{eq:metric}, we obtain
\begin{equation}
  g_{\mu \nu} = \sum_{0 < k < \pi} \frac{1}{64 E_k^2} (\partial_\mu \theta_k) (\partial_\nu \theta_k).
\end{equation}
With $\lambda_1 = J_1$ and $\lambda_2 = J_2$, this result may be written in matrix form as
\begin{align}
  g &= \sum_{0 < k < \pi} \frac{\sin^2 k}{64 E_k^6}\\
  &\times \begin{pmatrix}
    (h - J_2)^2 & (h - J_2) (2 h \cos k + J_1)\\
    (h - J_2) (2 h \cos k + J_1) & (2 h \cos k + J_1)^2
  \end{pmatrix}\nonumber.
\end{align}
In the thermodynamic limit, the momentum sum turns into an integral, which can be evaluated analytically. By expanding around the tricritical point, we obtain \eqref{eq:G}. 

\subsection{Adiabatic paths\label{app:ising_paths}}
\Tabref{tab:ising_paths} gives an explicit parametrization of the paths (i)--(iv) in \figref{fig:fig2}(a). The parameter $s$ ranges from $0$ to $1$. For path (ii), $s$ is related to $\beta$ by $s = \tanh \beta$.

\begin{table}[h]
  \centering
  \caption{\label{tab:ising_paths}}
  \begin{ruledtabular}
    \begin{tabular}{lll}
      path & $J_1/h$ & $J_2/h$\\
      \hline
      (i) & $2 s$ & $s$\\
      (ii) & $2 s$ & $s^2$\\
      (iii) & $3 (1 -s )^2 s + 7.5 (1-s) s^2$ & $1.5 (1 - s) s^2 + s^3$\\
      & \hfill $ + 2 s^3$ & \\
      (iv) & $6 (1 - s)^2 s + 9 (1-s) s^2$ & $-3 (1-s)^2 s + 4.5 (1-s) s^2 $\\
      & \hfill $+ 2 s^3$ & \hfill $+s^3$
    \end{tabular}
  \end{ruledtabular}
\end{table}

\section{Sampling from weighted independent sets}
\subsection{Numerical details\label{app:wis_numerical}}
To integrate the Schrödinger equation, we exactly diagonalize the Hamiltonian at discrete time steps $\Delta t$. The time steps are chosen such that $\Delta t || \di \ket{\text{0}} / \di t || = 10^{-3}$, where $\ket{\text{0}}$ denotes the instantaneous ground state. The expression is most conveniently evaluated using the identity $|| \di \ket{0} / \di t ||^2 = \sum_{n>0} |\bra{n} \di H / \di t \ket{0}|^2 / (E_n - E_0)^2$, where the sum runs over all excited states $\ket{n}$ with energy $E_n$.  The time is related to the parameters of the Hamiltonian by \eqref{eq:rate}.

\subsection{Adiabatic paths\label{app:wis_paths}}
\Tabref{tab:ising_paths} gives an explicit parametrization of the paths (i) and (ii) in \figref{fig:fig6}(a). The parameter $s$ ranges from $0$ to $1$. Along path (i), $s$ and $\beta$ are related by $s = e^{-\beta /2}$.
\begin{table}[ht]
  \centering
  \caption{\label{tab:wis_paths}}
  \begin{ruledtabular}
    \begin{tabular}{lll}
      path & $\Omega/V_g$ & $V_e/V_g$ \\
      \hline
      (i) & $s$ & $s^2$\\
      (ii) & $s$ & $6 s^2 - 5 s$\\
    \end{tabular}
  \end{ruledtabular}
\end{table}

\section{Two-state model for the star graph\label{app:star}}
The star graph has three types of vertices: the vertex at the center and the inner and outer vertices on each branch. If we maintain the permutation symmetry between the branches, the parent Hamiltonian takes the general form
\begin{align}
  H_q &= V_{e,\mathrm{cen}} n_\mathrm{cen} + V_{g,\mathrm{cen}} P_\mathrm{cen} (1 - n_\mathrm{cen}) - \Omega_\mathrm{cen} P_\mathrm{cen} \sigma_\mathrm{cen}^x \nonumber\\
  &+V_{e,\mathrm{in}} \sum_{i = 1}^b n_{\mathrm{in},i} + V_{g, \mathrm{in}} \sum_{i = 1}^b P_{\mathrm{in}, i} (1 - n_{\mathrm{in}, i}) \nonumber\\
  & \hspace{4.5cm}- \Omega_\mathrm{in} \sum_{i = 1}^b P_{\mathrm{in}, i} \sigma_{\mathrm{in}, i}^x \nonumber\\
  &+V_{e,\mathrm{out}} \sum_{i = 1}^b n_{\mathrm{out},i} + V_{g, \mathrm{out}} \sum_{i = 1}^b P_{\mathrm{out}, i} (1 - n_{\mathrm{out}, i}) \nonumber\\
  & \hspace{3.5cm}- \Omega_\mathrm{out} \sum_{i = 1}^b P_{\mathrm{out}, i} \sigma_{\mathrm{out}, i}^x,
\end{align}
where each row relates to a separate type of vertex and the sums run over all branches. With the weights specified in the main text, we have $V_{e, \mathrm{cen}} = e^{- b \beta}$, $V_{e, \mathrm{in}} = V_{e, \mathrm{out}} = e^{- \beta}$, $V_{g, \mathrm{cen}} = V_{g, \mathrm{in}} = V_{g, \mathrm{out}} = 1$, $\Omega_\mathrm{cen} = e^{- b \beta/2}$, $\Omega_{\mathrm{in}} = \Omega_\mathrm{out} = e^{- \beta/2}$ along the one-parameter family $H_q(\beta)$.

We restrict our analysis to the subspace that is completely symmetric under permutations of the branches. We introduce the total occupation numbers $n_\mathrm{in} = \sum_{i=1}^b n_{\mathrm{in}, i}$ and $n_\mathrm{out} = \sum_{i = 1}^b n_{\mathrm{out}, i}$ as well as the number of unoccupied branches $n_0$. The symmetric subspace is spanned by the states
\begin{equation}
  \ket{n_\mathrm{cen}, n_\mathrm{in}, n_\mathrm{out}, n_0},
  \label{eq:basis}
\end{equation}
where $n_\mathrm{cen} \in \{ 0, 1\}$ while the other occupation numbers are nonnegative integers satisfying $n_\mathrm{in} + n_\mathrm{out} + n_0 = b$. If $n_\mathrm{cen} = 1$, the independent set constraint further requires $n_\mathrm{in} = 0$. Each of the states in \eqref{eq:basis} is an equal superposition of $b!/ (n_\mathrm{in}! \, n_\mathrm{out}! \, n_0!)$ independent configurations. The dimension of the completely symmetric subspace is $(b + 1)(b+4)/2$.

The permutation symmetry leads to a bosonic algebra. We define the bosonic annihilation operators $b_\mathrm{in}$, $b_\mathrm{out}$, and $b_0$, respectively associated with the occupation numbers $n_\mathrm{in}$, $n_\mathrm{out}$, and $n_0$, which may be viewed as a generalization of Schwinger bosons. We split up the Hamiltonian into blocks where the central spin is either $0$ or $1$ as well as an off-diagonal term coupling them. Explicitly,
\begin{equation}
  H_q = H_q^{(0)} \otimes (1 - n_\mathrm{cen}) + H_q^{(1)} \otimes n_\mathrm{cen} + H_q^{(\mathrm{od})} \otimes \sigma_\mathrm{cen}^x.
\end{equation}
In terms of the bosonic operators
\begin{align}
  \label{eq:h0}
  H_q^{(0)} &= V_{g, \mathrm{cen}} P(n_\mathrm{in} = 0) + \\
  &\hspace{-.75cm} \begin{pmatrix}
    b_\mathrm{in}^\dagger, & b_\mathrm{out}^\dagger, & b_0^\dagger
  \end{pmatrix}
  \begin{pmatrix}
    V_{e, \mathrm{in}} & 0 & -\Omega_\mathrm{in}\\
    0 & V_{e, \mathrm{out}} & - \Omega_\mathrm{out}\\
    -\Omega_\mathrm{in} & -\Omega_\mathrm{out} & V_{g, \mathrm{in}} + V_{g, \mathrm{out}}
  \end{pmatrix}
  \begin{pmatrix}
    b_\mathrm{in}\\
    b_\mathrm{out}\\
    b_0
  \end{pmatrix},\nonumber\\
  \label{eq:h1}
  H_q^{(1)} &= V_{e, \mathrm{cen}} +
  \begin{pmatrix}
    b_\mathrm{out}^\dagger, & b_0^\dagger
  \end{pmatrix}
  \begin{pmatrix}
    V_{e, \mathrm{out}} & - \Omega_\mathrm{out}\\
    -\Omega_\mathrm{out} & V_{g, \mathrm{out}}
  \end{pmatrix}
  \begin{pmatrix}
    b_\mathrm{out}\\
    b_0
  \end{pmatrix} ,\\
  H_q^{(\mathrm{od})} &= - \Omega_\mathrm{cen} P(n_{\mathrm{in}} = 0),
\end{align}
where $P(n_\mathrm{in} = 0)$ projects onto states with no occupied inner vertices.

We diagonalize the Hamiltonian by treating the projectors perturbatively. We focus on the situation where all parameters follow the one-parameter family $H_q(\beta)$ except for $\Omega_\mathrm{cen}$ and $V_{e,\mathrm{cen}}$, which may be adjusted freely. By diagonalizing the matrices in \eqref{eq:h0} and \eqref{eq:h1}, we identify the lowest energy modes of the quadratic parts of $H_q^{(0)}$ and $H_q^{(1)}$ and associate with them the bosonic annihilation operators $c_0$ and $c_1$, respectively. Both modes have zero energy while the other modes are gapped at any finite value of $\beta$. We may thus expect the ground state to be well approximated in the subspace spanned by
\begin{equation}
  \ket{\psi_0} = \frac{1}{\sqrt{b!}} c_0^{\dagger b} \ket{\mathrm{vac}}, \qquad
  \ket{\psi_1} = \frac{1}{\sqrt{b!}} c_1^{\dagger b} \ket{\mathrm{vac}},
\end{equation}
where $\ket{\mathrm{vac}}$ denotes the bosonic vacuum. One can show that these states correspond to the Gibbs state of the star with the central spin held fixed.

Next, we perform a Schrieffer--Wolff transformation~\cite{Bravyi2011} to project onto the subspace spanned by $\ket{\psi_0}$ and $\ket{\psi_1}$. We arrive at an effective Hamiltonian
\begin{equation}
  H_\mathrm{eff} = 
  \begin{pmatrix}
    \varepsilon_0 + \delta \varepsilon_0 & -J - \delta J\\
    -J - \delta J & V_{e,\mathrm{cen}} + \delta \varepsilon_1
  \end{pmatrix},
\end{equation}
where the terms
\begin{gather}
  \varepsilon_0 = \bra{\psi_0} P(n_\mathrm{in} = 0) \ket{\psi_0} = \left( \frac{1 + e^{-\beta}}{2 + e^{-\beta}} \right)^b, \\
  J = \Omega_\mathrm{cen} \bra{\psi_1} P(n_\mathrm{in} = 0) \ket{\psi_0} = \Omega_\mathrm{cen}\left( \frac{1 + e^{-\beta}}{2 + e^{-\beta}} \right)^{b/2}
\end{gather}
are obtained by projecting the full Hamiltonian onto the low-energy subspace. The correction from coupling to excited states, as given by the Schrieffer--Wolff transformation to lowest nontrivial order, are 
\begin{align}
  \delta \varepsilon_0 &= - \varepsilon_0 \sum_{n} \frac{1}{E_n } \left| \bra{n}  \sigma_\mathrm{cen}^x \ket{\psi_1} \right|^2 \\
  \delta \varepsilon_1 &= - \Omega_\mathrm{cen}^2 \sum_{n} \frac{1}{E_n} \left| \bra{n}  \sigma_\mathrm{cen}^x \ket{\psi_1} \right|^2 \\
  \delta J &= -\Omega_\mathrm{cen} \sqrt{\varepsilon_0} \sum_{n} \frac{1}{E_n} \left| \bra{n}  \sigma_\mathrm{cen}^x \ket{\psi_1} \right|^2
\end{align}
where we used the relation $P(n_\mathrm{in} = 0) \ket{\psi_0} = \sqrt{\varepsilon_0} \sigma_\mathrm{cen}^x \ket{\psi_1}$, which holds along the paths of interest. The sums run over all excited states $\ket{n}$ with energy $E_n$ of the unperturbed part of $H_q^{(0)}$. We neglected a term $V_{e,\mathrm{cen}}$ in the energy denominator, which is justified as long as $V_{e,\mathrm{cen}}$ is small compared to $E_n$. The discussion remains valid even if this is not the case because the shifts from the Schrieffer--Wolff transformation can then be ignored as far as the ground state is concerned.

The complete effective Hamiltonian may be written as
\begin{equation}
  H_\mathrm{eff} = 
  \begin{pmatrix}
    (1 - f) \varepsilon_0 & -( 1 - f) J\\
    -(1- f) J & V_{e,\mathrm{cen}} - f \Omega_\mathrm{cen}^2
  \end{pmatrix},
\end{equation}
where $f = \sum_n \left| \bra{n}  \sigma_\mathrm{cen}^x \ket{\psi_1} \right|^2 / E_n$. We find numerically that $f$ decays as an inverse power law in $b$ such that our approximations are well justified in the thermodynamic limit. Along the one-parameter family $H_q(\beta)$, we have $V_{e, \mathrm{cen}} = \Omega_\mathrm{cen}^2$ such that $H_\mathrm{eff}$ depends on $f$ only through an overall factor $(1 - f)$, which tends to $1$ in the limit of large $b$.

\bibliography{bibliography}
\end{document}